\documentclass[aps,prd,superscriptaddress,showpacs,nofootinbib,notitlepage,twocolumn]{revtex4-1} 

\usepackage{graphicx}
\usepackage{bm}
\usepackage{dcolumn}
\usepackage{amssymb}
\usepackage{amsmath}
\usepackage{epsfig}
\usepackage[english]{babel}
\usepackage[utf8]{inputenc}
\usepackage{eucal}
\usepackage{hyperref}
\usepackage{verbatim}
\usepackage{latexsym}
\usepackage{color}
\usepackage{slashed}
\usepackage{ulem}

\DeclareMathOperator{\Tr}{Tr}
\DeclareMathOperator{\Det}{Det}

\begin{document}

\title{Chiral and deconfinement Transitions in spin-polarized quark matter}

\author{Ricardo L. S. Farias} 
\affiliation{Departamento de F\'{\i}sica, Universidade Federal de Santa Maria, 
97105-900, Santa Maria, RS, Brazil} 
\email{ricardo.farias@ufsm.br}

\author{William R. Tavares}
\affiliation{Departamento de F\'{\i}sica Te\'orica, Universidade do
  Estado do Rio de Janeiro, 20550-013 Rio de Janeiro, RJ, Brazil}  
\email{tavares.william@ce.uerj.br}

\begin{abstract}
We investigate the influence of spin polarization in strongly interacting matter by introducing a finite spin potential, $\mu_\Sigma$, which effectively controls the spin density of the system without requiring rotation or specific boundary conditions. Inspired by recent lattice QCD simulations that incorporated such a potential, we implement this approach within an effective QCD framework. Our results show that increasing spin polarization leads to a simultaneous decrease in both the chiral and deconfinement restoration temperatures. The resulting phase structure is qualitatively consistent with lattice findings, and notably, we observe the emergence of a first-order chiral phase transition at low temperature. These results suggest that spin-polarized environments can significantly impact the QCD phase diagram and offer a controlled route for studying spin effects in hot and dense matter.
\end{abstract}

\maketitle

\section{Introduction}
The quark-gluon plasma is not only an almost perfect fluid~\cite{Busza:2018rrf}, but also the most vortical, as the detection of the global polarization of $\bar{\Lambda}$ and $\Lambda$ baryons~\cite{STAR:2017ckg} indicates. The estimated values of angular velocity associated with the system can be predicted by experimental constraints as $\omega\sim 7$ MeV~\cite{STAR:2017ckg}, or obtained with hydrodynamic models as $\omega\sim 20 - 40$ MeV~\cite{Jiang:2016woz}. The underlying physics of these new effects is still under investigation, and some experiments conducted by the ALICE and STAR Collaborations indicate, for instance, different values of the spin-density matrix $\rho_{00}$~\cite{ALICE:2019aid,STAR:2022fan}. Various groups engaged in lattice quantum chromodynamics (LQCD) and effective model studies have been exploring different aspects of the QCD phase diagram~\cite{Braguta:2021ucr,Braguta:2021jgn,Braguta:2022str,Yang:2023vsw,Braguta:2023aio,Chernodub:2022veq,Jiang:2016wvv,Chernodub:2016kxh,Sun:2021hxo,Sun:2023yux,Wang:2018sur,Singha:2024tpo,Hernandez:2024nev}, introducing also new effects such as the negative Barnett effect~\cite{Braguta:2023tqz,Braguta:2023yjn,Braguta:2023qex} and the chiral vortical catalysis~\cite{Jiang:2021izj,Nunes:2024hzy}. 

The theoretical problem, in general, is characterized by strongly interacting matter in a finite-sized rigid cylinder with constant angular velocity $\omega$, in which one needs to set $\omega r \leq 1$, where $r$ is the radius of the cylinder, in order to obey causality~\cite{Chernodub:2016kxh}. Also, several applications focus on homogeneous chiral condensates valid for small angular velocities~\cite{Chernodub:2016kxh,Chen:2021aiq,Singha:2024tpo}, while nonhomogeneous phases have been predicted recently in different applications~\cite{Chernodub:2020qah,Chernodub:2022veq,Braguta:2023iyx}. Effective approaches, e.g., the Nambu--Jona-Lasinio model~\cite{Jiang:2016wvv,Chernodub:2016kxh,Wang:2018sur,Sun:2023yux,Sun:2024anu,Sun:2021hxo,Ghalati:2023npr,Sadooghi:2021upd,Hua:2024bwn,Chen:2024utf,Sun:2023kuu,TabatabaeeMehr:2023tpt,Wei:2023pdf,Chen:2021aiq,Mehr:2022tfq,wei2022mass,Chen:2019tcp,Zhang:2018ome,Cao:2023olg}, the linear sigma model coupled with quarks~\cite{Singha:2024tpo,Hernandez:2024nev, Chernodub:2024wis,Chen:2023cjt,Wan:2020ffv,Chen:2020ath} and holographic approaches~\cite{Chen:2024jet,Braga:2023qej,Braga:2023qee,Zhao:2022uxc,Yadav:2022qcl}, have been applied in order to predict possible phase diagrams, i.e., pseudocritical temperature as a function of the angular velocity, $T_{pc}\times\omega$, that can be dependent on different boundary conditions associated with fermionic degrees of freedom in a cylindrical system \cite{Chernodub:2016kxh,Chen:2021aiq,Ebihara:2016fwa}. In LQCD the phase diagram indicates an increasing behavior of $T_{pc}\times\omega$~\cite{Braguta:2021jgn,Braguta:2022str}, while effective approaches predict the inverse behavior, which is associated with the lack of gluons in these methods~\cite{Jiang:2021izj,Nunes:2024hzy,Sun:2024anu}. 

One of the important aspects of vorticity in strongly interacting matter is the spin polarization of the particles in the system. The role of spin polarization has instigated studies in different contexts, e.g., relativistic fluid dynamics and effective field theories \cite{Montenegro:2017rbu,Florkowski:2017ruc} A possible approach to introduce this condition effectively, without dealing with boundary conditions associated with the geometry of the system, is to apply a finite spin potential, $\mu_{\Sigma}$, as recently proposed in Ref.~\cite{Braguta:2025ddq} using two-flavor Monte Carlo simulations. However, the generalized spin-tensor contribution suffers from ambiguities in relativistic spin-hydrodynamics associated with pseudogauge transformations~\cite{Fukushima:2020ucl,Buzzegoli:2024mra,Speranza:2020ilk,Becattini:2018duy,Fang:2025aig,Huang:2024ffg}, which can be fixed by interacting field theories~\cite{Buzzegoli:2024mra}. Then, a simplification of the spin-density matrix term can be applied in order to choose a preferred polarization in the $z$ direction. Physically, in the same way we think about the finite baryon density as a conjugated variable of baryon chemical potential, one can expect a spin density associated with a spin potential. This prescription is therefore interesting for studying a system composed by fermions with a specific density of polarized spins in a chiral symmetry broken environment without invoking rotations and avoiding issues of the thermodynamic limit~\cite{Braguta:2025ddq}. An additional advantage is that the spin-potential is applied, at this initial point, only to the fermionic degrees of freedom and is well-suited for effective models and presents a possibility to make reliable comparisons with LQCD.

In this work, we apply the spin-potential control parameter in the $z$ direction to the two-flavor entangled Polyakov--Nambu--Jona-Lasinio (EPNJL) model~\cite{Sakai:2010rp,Sakai:2009dv} in the mean-field approximation. One of the benefits of using an effective model is to avoid the sign problem observed in LQCD~\cite{Yamamoto:2011gk}, which allows us to explore real values of $\mu_{\Sigma}$ in all regions of interest. In this way, the main idea of this manuscript is to compare recent lattice results with the ones obtained by the model and investigate some of the basic predictions not reached by LQCD. To this end, we obtain the solution of the dispersion relation of quarks associated with the inclusion of $\mu_{\Sigma}$. Then, we include the Polyakov loop potential with the entanglement interaction in the thermodynamical potential in order to match the pseudocritical temperatures of chiral and deconfinement phase transitions with lattice at $\mu_{\Sigma}=0$. We calculate the effective quark masses and the expectation value of the Polyakov loop at finite values of temperature and spin-density potential. By looking at the peak of the susceptibilities we obtain the pseudocritical temperatures. The dependence $T_{pc}(\mu_{\Sigma})$ is explored by applying a quadratic fitting function of $\mu_{\Sigma}$ and looking at the curvature $\kappa$ and comparing it with LQCD. We also predict how the effective quark masses behave at higher values of the spin-density potential at low temperature.

The manuscript is structured as follows. The basic steps to obtain the quark dispersion relation of the SU(2) NJL model with the spin-potential are shown in Sec.~\ref{muSigma}. In Sec.~\ref{sec1} we present the model details of the SU(2) EPNJL. In Sec.~\ref{NR} we present and discuss our results. The conclusions are shown in Sec.~\ref{conclusions}.

\section{Finite spin potential}
\label{muSigma}

In order to show the quark dispersion relation with the inclusion of the spin-potential, the finite spin density is introduced
by the quark spin potential in the canonical formulation of the spin operator, following the approach developed in~\cite{Braguta:2025ddq}. The spin potential exhibits a distant resemblance to both axial and helical chemical potentials. Notably, at zero temperature, we observe an intriguing analogy between systems governed by an imaginary quark spin potential 
$\mu_{\Sigma}^I$	
  in Euclidean spacetime and those governed by an axial (chiral) chemical potential 
$\mu_A$, in Minkowski spacetime~\cite{Ambrus:2019khr,Chernodub:2020yaf,Braguta:2025ddq,Brandt:2024wlw}.

We solve the generating functional for the SU(2) Nambu--Jona-Lasinio and the extension to include the Polyakov loop and  entanglement is straightforward~\cite{Sakai:2010rp} and will be explored in the next section. 

The general Lagrangian with spin-potential is given by

\begin{eqnarray}
\mathcal{L}=\overline{\psi}\left(i \slashed \partial - \hat{m} \right)\psi
+G\left[(\overline{\psi}\psi)^{2}+(\overline{\psi}i\gamma_{5}\vec{\tau}\psi)^{2}\right]+\mathcal{L}_{\Sigma},~\label{su2njl}
\end{eqnarray}

\noindent where $G$ is the coupling 
constant and $\vec{\tau}$ are isospin Pauli matrices; the bare quark mass matrix $\hat{m}=$diag($m_u$, $m_d$) and  
 $\psi=(\psi_u \quad \psi_d)^T$ is the quark field. We consider the isospin limit $m_u$=$m_d=m_c$. Regarding the quark spin density term applied to the NJL model, we have 

\begin{eqnarray}
\mathcal{L}_{\Sigma}=\mu_{\alpha,\mu\nu}\bar{\psi}S^{\alpha,\mu\nu}\psi.
\end{eqnarray}

\noindent where the following definitions are used:

\begin{eqnarray}
    S^{\alpha,\mu\nu}=\frac{1}{2}\{\ \gamma^{\alpha},\Sigma^{\mu\nu}\}\ ,\quad \Sigma^{\mu\nu}=\frac{i}{4}[\gamma^{\mu},\gamma^{\nu}].
\end{eqnarray}

Since we are interested in the polarization in the $z$ direction we set  

\begin{eqnarray}
    \mu_{\alpha,\mu\nu}=\frac{\mu_{\Sigma}}{2}\delta_{\alpha0}(\delta_{\mu1}\delta_{\nu2}-\delta_{\nu1}\delta_{\mu2}).
\end{eqnarray}

With this choice, the final Lagrangian becomes

\begin{eqnarray}
\mathcal{L}&&=\overline{\psi}\left(i \slashed \partial - m_c + \frac{\mu_\Sigma}{2}\gamma^3\gamma^5\right)\psi
+G\left[(\overline{\psi}\psi)^{2}\right.\nonumber\\
&&+\left.(\overline{\psi}i\gamma_{5}\vec{\tau}\psi)^{2}\right].
\label{su2njl}
\end{eqnarray}

 The spin polarization in the system is controlled by the spin-potential $\mu_\Sigma$, as originally developed in Ref.~\cite{Braguta:2025ddq}. We use the Dirac representation for the $\gamma$ matrices. 
 
To find the quark dispersion relation we start by looking to the functional generator:

 \begin{align}
     Z=\mathcal{N}\int D\bar{\psi}D\psi \exp\left[{i\int d^4x\mathcal{L}(\bar{\psi},\psi)}\right],
 \end{align}

\noindent where $\mathcal{N}$ is a normalization constant. We then perform the Hubbard-Stratanovich transformation to introduce auxiliary fields $\sigma$ and $\vec{\pi}$, which are defined as 

\begin{eqnarray}
\vec{\pi}=-2G\bar{\psi}i\gamma_5\vec{\tau}\psi,\quad\sigma = -2G\bar{\psi}\psi.
\end{eqnarray}

Assuming an isospin-symmetric system, i.e., $\vec{\pi}=0$, the partition function becomes
$Z=\mathcal{N}Z_1 Z_2$, where:

\begin{eqnarray}
&&Z_1=\exp\left[{-i\frac{\sigma^2}{4G}}\int d^4x \right],\\
&& Z_2=\int D\bar{\psi}D\psi \exp\left[i\int d^4x \bar{\psi}(i\slashed{\partial}-m_c-\sigma \right. \nonumber\\
&&+\left.\frac{\mu_{\Sigma}}{2}\gamma_3\gamma_5)\psi\right].
\end{eqnarray}

After performing the functional integration over the Grassmann variables:

\begin{eqnarray}
Z_2 &&= \exp \left[N_c \int d^4x \int \frac{d^4p}{(2\pi)^4}\Tr_{f,D}\ln (\slashed{p}-m_c-\sigma \right.\nonumber\\
&&+\left.\frac{\mu_{\Sigma}}{2}\gamma_3\gamma_5)\right],
\end{eqnarray}

\noindent with $N_c=3$. Evaluating the trace over the flavor and Dirac indices using $\Tr \ln\mathcal{O} \equiv \ln \Det\mathcal{O}$ we obtain

\begin{eqnarray}
    Z_2&&=\exp
\left\{N_c\sum_{s=\pm1}\int d^4x \int \frac{d^4 p}{(2\pi)^4}\ln\left[-p_0+E_s(\vec{p})\right]^2\right.\nonumber\\
&&\times\left.\left[p_0+E_s(\vec{p})\right]^2\right\},
\end{eqnarray}

\noindent where the dispersion relation $E_p^s$ is given by

\begin{eqnarray}
    E_s=\sqrt{p_\perp^2 +\left(\sqrt{p_3^2+M^2}+s\frac{\mu_{\Sigma}}{2} \right)^2},
\end{eqnarray}

\noindent where $M=m+\sigma$ and $p_{\perp}^2=p_1^2+p_2^2$. 
  
\section{EPNJL model and equations}\label{sec1}

The generalized form of the Lagrangian of the PNJL model for $N_f = 2$ light quarks and $N_c = 3$
color degrees of freedom, where the quarks are coupled to a temporal background gauge 
field, represented in terms of Polyakov loops is given by~\cite{Ratti:2005jh}

\begin{eqnarray}
\mathcal{L}=&&\overline{\psi}\left(i \slashed D - m_c + \frac{\mu_\Sigma}{2}\gamma^3\gamma^5 + \mu\gamma^0\right)\psi
+G\left[(\overline{\psi}\psi)^{2}\right.\nonumber\\
&&+\left.(\overline{\psi}i\gamma_{5}\vec{\tau}\psi)^{2}\right] -\mathcal{U}(l,\overline{l},T)\, .\label{su3pnjl}
\end{eqnarray}

\noindent Here, the covariant derivative is given by $D^\mu =(i\partial^{\mu}-i\mathcal{A}^{\mu})$ 
 where $\mathcal{A}^{\mu}=\delta^{\mu}_0\mathcal{A}^0$ is the Polyakov gauge. The strong coupling constant $g$ is absorbed into the definition $\mathcal{A}^{\mu}(x)=g\frac{\lambda_a}{2}\mathcal{A}_a^{\mu}(x)$,
$\lambda_a$ are the Gell-Mann matrices and $\mathcal{A}_a^{\mu}(x)$ is the $SU(3)$ gauge field. 

For the pure gauge sector, let us define the Polyakov line as
\begin{eqnarray}
 L(x)=\mathcal{P}\exp\left[i\int_0^{\beta}d\tau\mathcal{A}_4(\tau,\overrightarrow {x}) \right],
\end{eqnarray}
\noindent where $\mathcal{P}$ denotes path ordering and $\beta=\frac{1}{T}$. Also, $\mathcal{A}_4=i\mathcal{A}_0$ is the temporal 
component of the Euclidean gauge field 
$(\mathcal{A}_4,\overrightarrow{\mathcal{A}})$.

The effective potential $\mathcal{U}(l,\overline{l},T)$ for the Polyakov fields is parametrized in 
order to reproduce lattice results 
in the mean-field 
approximation~\cite{Ratti:2006wg,Ratti:2005jh}. We adopt the ansatz~\cite{Ratti:2005jh}, 

\small{
\begin{align}
\frac{\mathcal{U}(l,\overline{l},T)}{T^4}=-\frac{a(T)}{2}l\overline{l} +b(T)\ln\left[1-6l\bar{l}+4(l^3+\overline{l}^3)-3(l\bar{l})^2\right], \label{potPNJL}
\end{align}
}

\noindent where $a(T)$ and $b(T)$ are given by

\begin{eqnarray}
 a(T)=a_0+a_1\left(\frac{T_0}{T}\right)+a_2\left(\frac{T_0}{T}\right)^2,\quad b(T)=b_3\left(\frac{T_0}{T}\right)^3.
\end{eqnarray}

The parameters of the potential $\mathcal{U}(l,\overline{l},T)$ 
will be specified in the numerical results section. 
We should also mention that the thermal expectation value of the Polyakov loop is given by~\cite{Fukushima:2013rx}

\begin{eqnarray}
l \equiv \frac{1}{N_c} \left \langle Tr_c L(x) \right \rangle, \quad \overline{l} 
\equiv \frac{1}{N_c} \left \langle Tr_c L^\dagger(x) \right \rangle .
\end{eqnarray}

Since we are working with the background field $\mathcal{A}_4$, we can 
 obtain the generalization of all physical quantities at finite 
temperature and density using
 the following symbolic replacements~\cite{Hansen:2006ee} 

\begin{eqnarray}
 i\int\frac{d^4p}{(2\pi)^4}&\rightarrow&-T\frac{1}{N_c}\Tr_c\sum_{n=-\infty}^{\infty}\int\frac{d^3p}{(2\pi)^3},\label{intrep}\\
 (p_0,\overrightarrow{p})&\rightarrow& (i\omega_n+\mu-i\mathcal{A}_4,\overrightarrow{p}),\label{repl}
\end{eqnarray}

\noindent where $\omega_n=(2n+1)\pi T$ is the Matsubara frequency. 
 In this work, we consider only the zero baryon chemical potential case, and hence $l$=$\overline{l}$.

The inclusion of the entanglement is implemented by the change in the coupling

\begin{eqnarray}
    G\rightarrow G(l,\bar{l})=G\left[1-\alpha_1l\bar{l}-\alpha_2(l^3+\bar{l}^3)\right].
\end{eqnarray}

The final expression for the effective potential is given by 
\begin{eqnarray}
\Omega(M,l,l^{\dagger},T,\mu) & = & \mathcal{U}\left(l,l^{\dagger},T\right)
+\frac{\left(M-m_{0}\right)^{2}}{4G(l,\bar{l})} + \Omega_V \nonumber\\
&&-  \frac{N_{f}}{\beta}\sum_{s=\pm1}\int\frac{d^{3}p}{\left(2\pi\right)^{3}}
 \left[\log\left(F_{+}^{s}F_{-}^{s}\right)\right],
 \label{V1pnjl}
\end{eqnarray}
where $E_{s}^2(p_3,p_{\perp})=p_\perp^2  + \left(\sqrt{p_3^2+M^{2}} + s\frac{\mu_\Sigma}{2}\right)^2$, and we adopt the following definitions:

\begin{align}
& \Omega_V  =  - N_{c}N_{f}\sum_{s=\pm1}\int\frac{dp_3dp_{\perp}~p_{\perp}}{2\pi^2}E_{s}(p_3,p_{\perp}),\\
& F_{-}^{s}  =  1+3 l e^{-\beta\left(E_{s}-\mu\right)}
+3 l^{\dagger}e^{-2\beta\left(E_{s}-\mu\right)}+e^{-3\beta\left(E_{s}-\mu\right)},\\
& F_{+}^{s}  =  1+3 l^{\dagger}e^{-\beta\left(E_{s}+\mu\right)}
+3 l e^{-2\beta\left(E_{s}+\mu\right)}+e^{-3\beta\left(E_{s}+\mu\right)}\;.
\end{align}

By minimizing the thermodynamic potential, Eq. (\ref{V1pnjl}), with respect
to $M$, $l$, and $l^{\dagger}$, we obtain the three gap equations:

\begin{eqnarray}
\frac{\partial\Omega(M,l,l^{\dagger},T,\mu)}{\partial M} &=& 
\frac{\partial\Omega(M,l,l^{\dagger},T,\mu)}{\partial l} = 0, \nonumber\\ 
\frac{\partial\Omega(M,l,l^{\dagger},T,\mu)}{\partial l^{\dagger}} &=& 0.
\label{gaps}
\end{eqnarray}
Numerical results for these quantities will be presented in the next section.

\section{Numerical Results}
\label{NR}

In this section we show our numerical results. The basic parameter set for $T=0$ is chosen to be $\Lambda=587.9$ MeV, $G\Lambda^2=2.44$ and
$m_c=5.6$ MeV~\cite{Buballa:2003qv}. The parameters associated with the Polyakov loop potential are given in Table\ref{table1}. The pure gauge sector, the transition temperature is given by $T_0=270$ MeV~\cite{Sakai:2009dv}, however, a lower value of $T_0$ is usually necessary in order to include the effects of $N_f=2$ in the theory. Following the procedure adopted in references~\cite{Sakai:2009dv,Endrodi:2019whh,Schaefer:2007pw}, we use $T_0=212$ MeV. For the entanglement interaction we adopt $\alpha_1=\alpha_2=0.2$~
\cite{Sakai:2010rp}.

%Tab.~\ref{table1}
\begin{table}[ht!]
\caption{\label{table1} Parameters for EPNJL using logarithmic parametrization.}
\begin{center}
\begin{tabular}{ccccccc}
\hline 
\hspace*{.3cm}$a_{0}$\hspace*{.3cm} & \hspace*{.3cm}$a_{1}$ \hspace*{.3cm}
 &\hspace*{.3cm} $a_{2}$\hspace*{.3cm} &\hspace*{.3cm}
 $b_{3}$\hspace*{.3cm} &\hspace*{.3cm}
 $T{}_{0}$ [MeV]\\
\hline 
3.51 & -2.47 & 15.2  & -1.75  & 0.212  \\\hline 
\end{tabular}
\end{center}
\end{table}

In EPNJL model we have two different critical (pseudocritical) temperatures. The first
one is related to the chiral symmetry restoration, $T_{pc}^{c}$, 
and the second is the deconfinement temperature, $T_{pc}^d$, defined as 

\begin{equation}
 T_{pc}^{c}\to -\frac{\partial M}{\partial T}\;\;\;\;\;\;\;\;\;\text{and}
 \;\;\;\;\;\;\;\;\; T_{pc}^{d}\to \frac{\partial l}{\partial T}.
 \label{Tpc}
\end{equation}

Once both transitions are crossovers at high enough temperatures, we determine the critical values through the 
peaks of the quantities.

\begin{figure}[htb]
\vspace{0.75cm}
\includegraphics[scale=0.44]{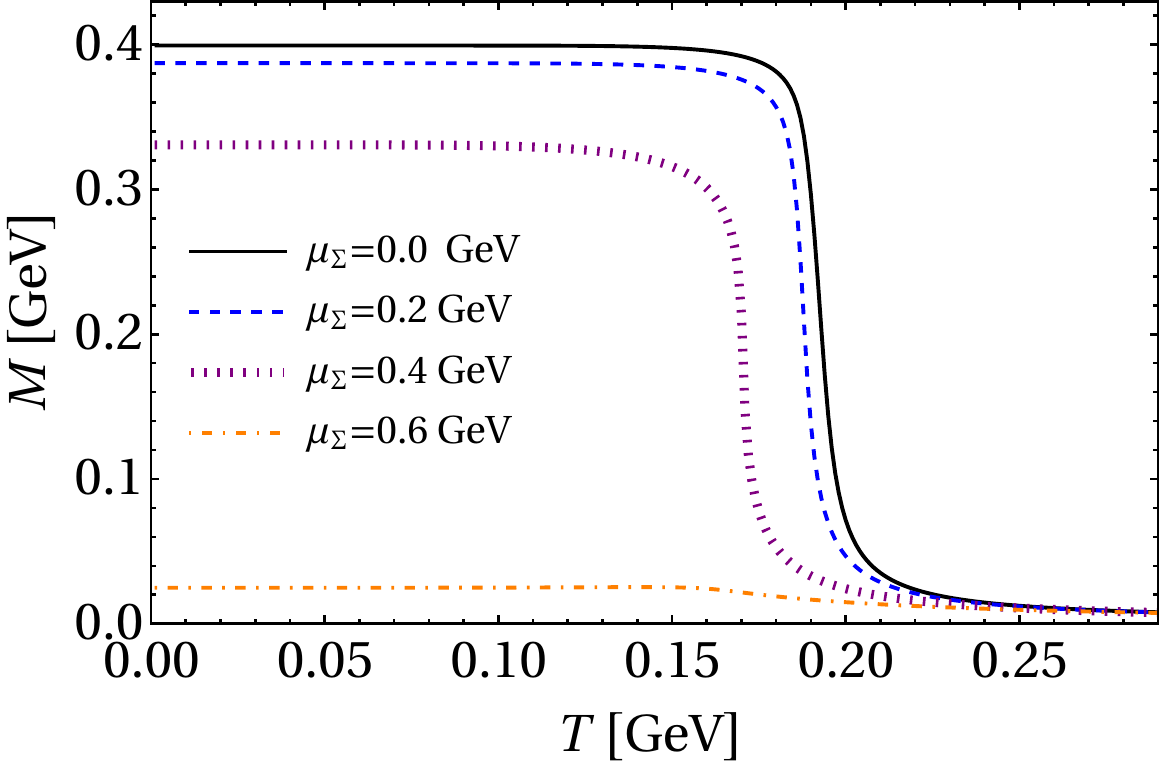}
\includegraphics[scale=0.44]{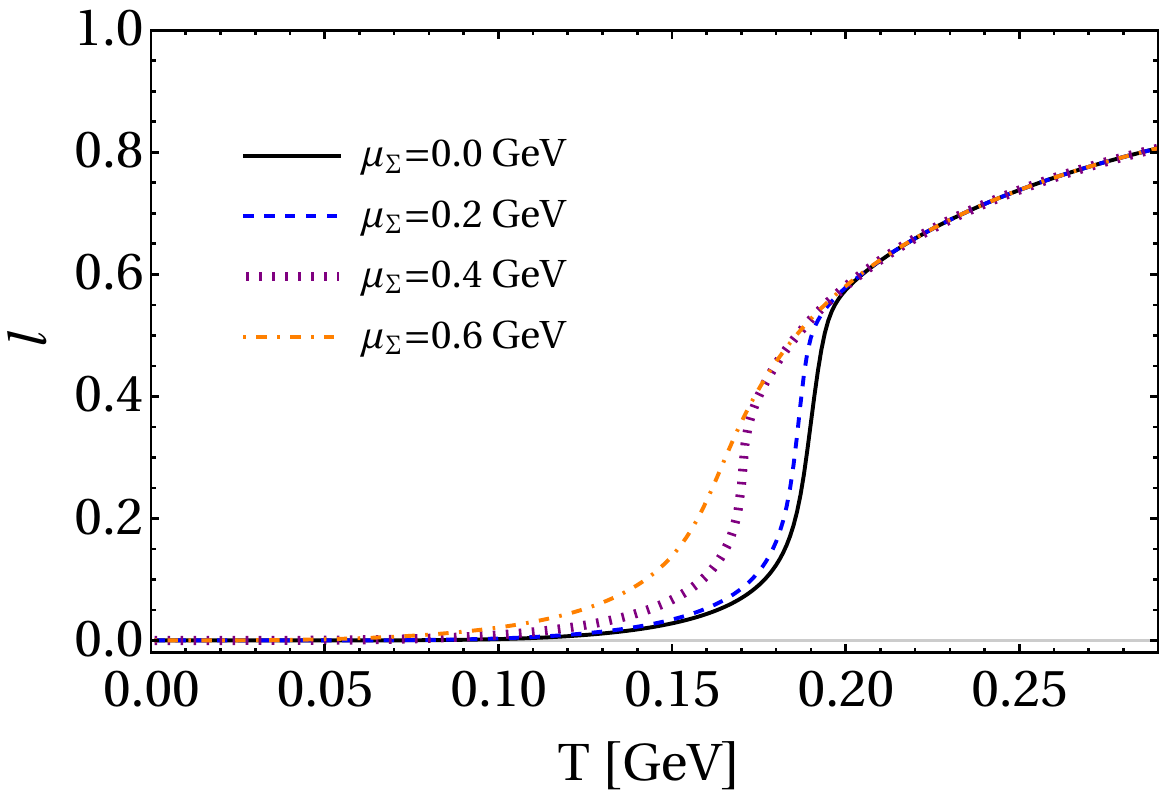}
\caption{Constituent quark mass (top panel) and the expected value of Polyakov loop (lower panel) as a function of the temperature for different values of $\mu_\Sigma$.}
\label{m_phi}
\end{figure}

In Fig. \ref{m_phi} we show the effective quark masses (top panel) and the expectation values of the Polyakov loop (lower panel) as functions of the temperature for different values of spin potential $\mu_{\Sigma}$. It is evident that, at $T=0$ the masses decrease as we increase $\mu_{\Sigma}$; the results also suggest a first-order phase transition at $\mu_{\Sigma}\sim 0.472$ GeV, in which an abrupt decrease is observed. As we increase the temperature the spin potential and temperature together enhance a partial chiral symmetry restoration, indicating a decreasing behavior of the pseudocritical temperature as $\mu_{\Sigma}$ increases. For the Polyakov loop expectation value, the transition for the deconfined phase shifts to lower temperatures.

%%%%%%%%%%%%%%%%%%%%%%%%%%%%%%%%%%%%%%%%%%5
%%%%%%%%%%%%%%%%%%%%%%%%%%%%%%%%%%%%%%%%%%%%

\begin{figure}[htb]
\vspace{0.75cm}
\includegraphics[scale=0.44]{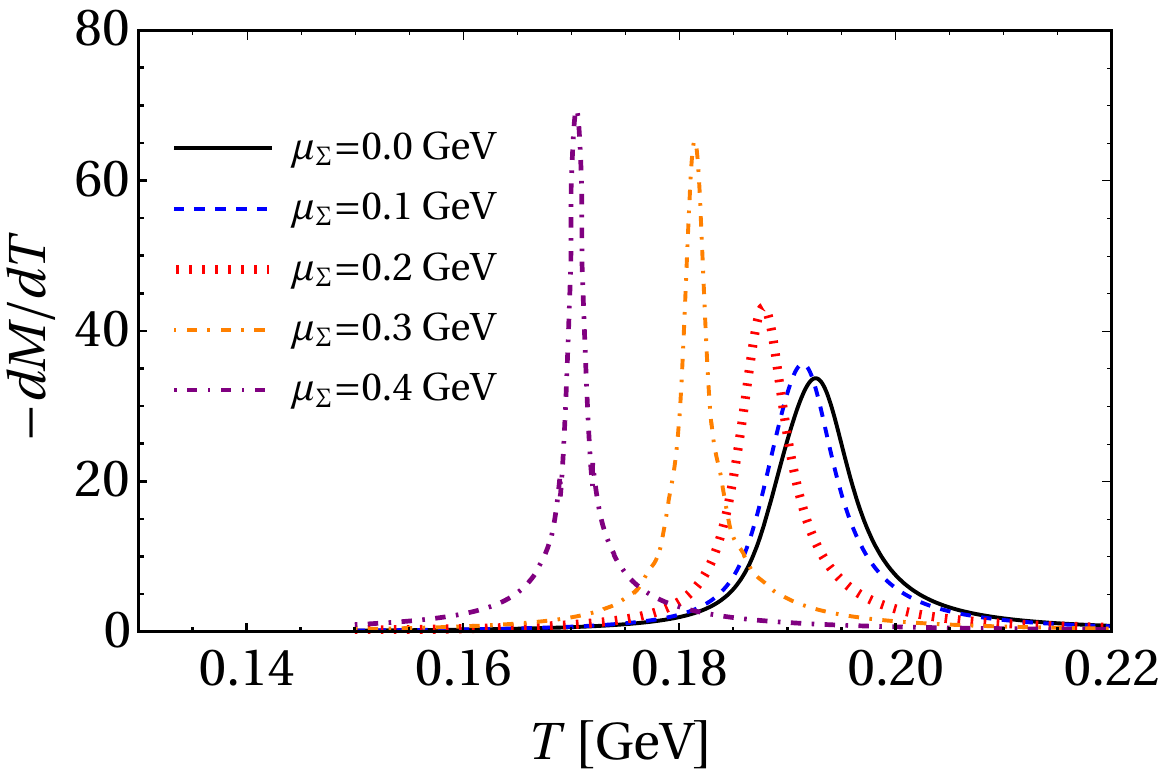}
\includegraphics[scale=0.44]{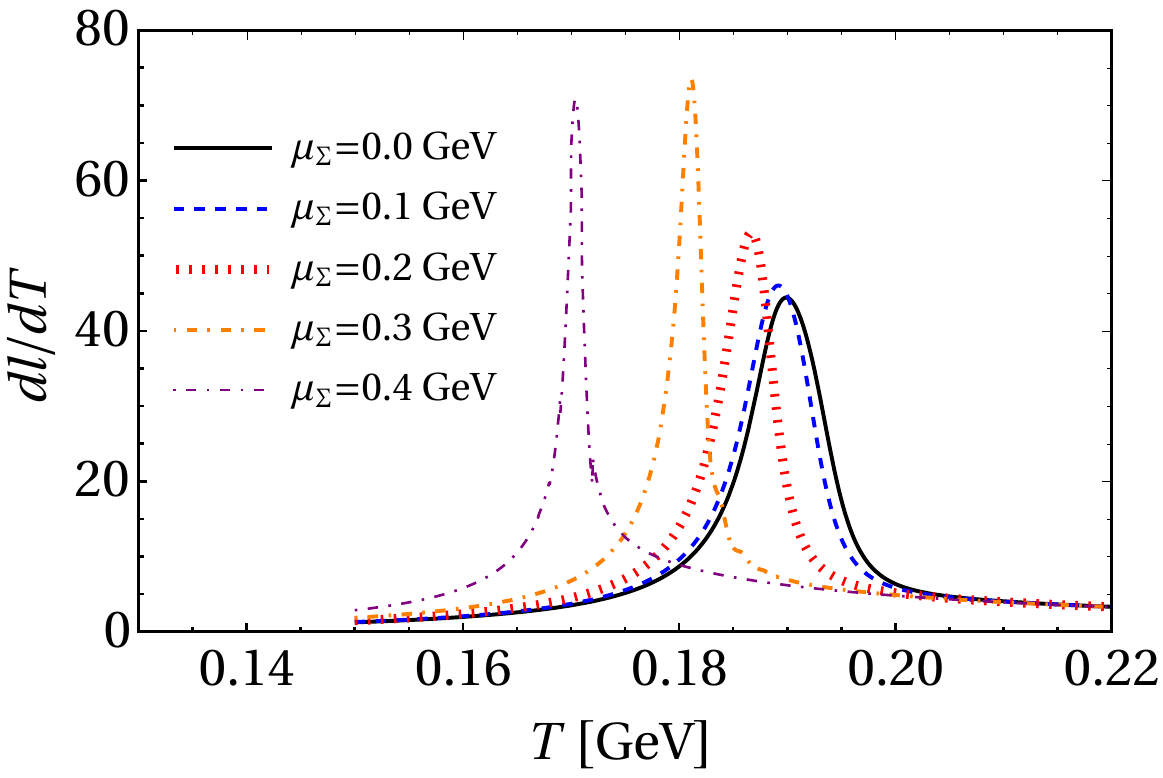}
\caption{Chiral and deconfinement susceptibilities as a function of temperature for different values of spin potential $\mu_\Sigma$.}
\label{susc}
\end{figure}

The susceptibilities of chiral and deconfinement phase transitions are shown in Fig.\ref{susc} as functions of the temperature for different values of $\mu_{\Sigma}$. As explained earlier, the peaks are treated as the pseudocritical temperatures of chiral and deconfinement phase transitions. It is clear that a decreasing behavior of both pseudocritical temperatures occurs as $\mu_{\Sigma}$ increases, which is in qualitative agreement with recent LQCD data~\cite{Braguta:2025ddq}. 

%%%%%%%%%%%%%%%%%%%%%%%%%%%%%%%%%%%%%%%%%5

%%%%%%%%%%%%%%%%%%%%%%%%%%%%%%%%%%%%%%%%%%

\begin{figure}[htb]
\vspace{0.75cm}
\includegraphics[scale=0.44]{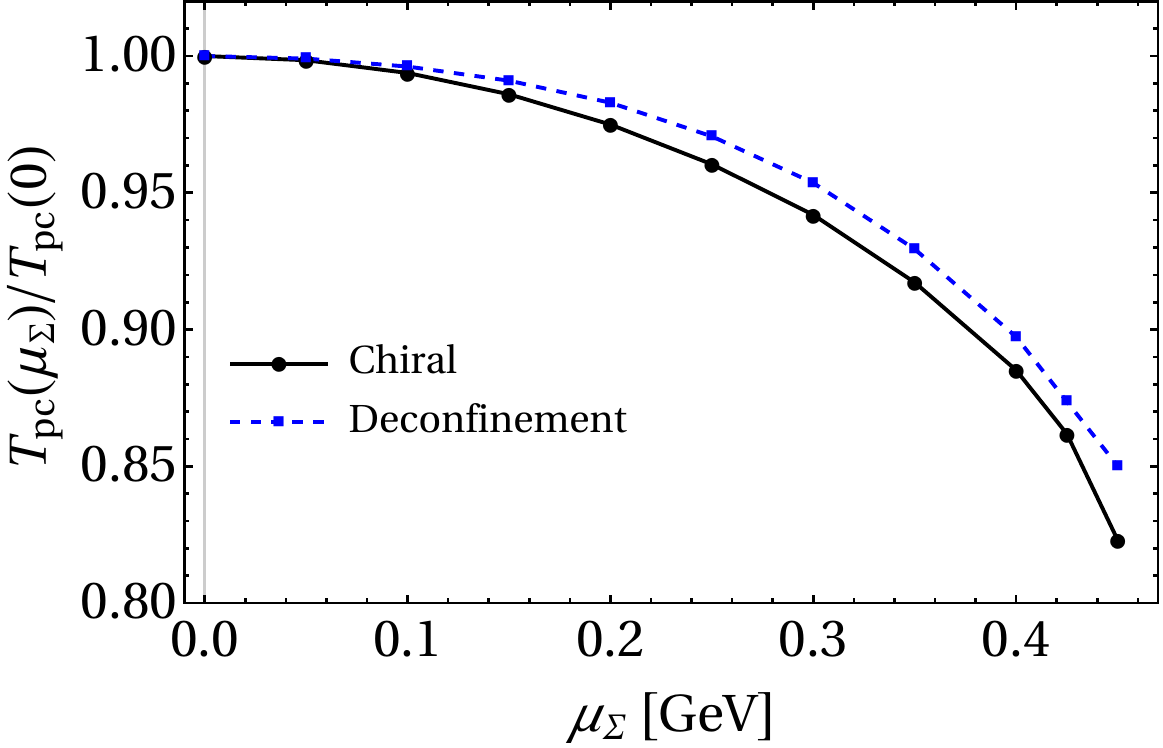}
\caption{Normalized pseudocritical temperatures of chiral and deconfinement phase transitions as a function of the spin potential $\mu_\Sigma$.}
\label{Txmu}
\end{figure}

Since we are using the EPNJL model, it is expected that both pseudocritical temperatures, $T_{pc}^d$ and $T_{pc}^c$, are very close, which is indeed confirmed in Fig. \ref{Txmu}. The application of the entangled extended EPNJL was motivated by the goal of reducing the splitting between both pseudocritical temperatures~\cite{Sakai:2010rp} and, in this way, being more aligned with the findings of LQCD~\cite{Braguta:2025ddq}. 

%%%%%%%%%%%%%%%%%%%%%%%%%%%%%%%%%%%%%%%%%%%%%%%%%

%%%%%%%%%%%%%%%%%%%%%%%%%%%%%%%%%%%%%%%%%%

 It is important to mention that, the LQCD results in Ref.~\cite{Braguta:2025ddq} found a fitting function for the pseudocritical temperature as a function of $\mu_{\Sigma}$ by the quadratic function

\begin{eqnarray}
\frac{T_{pc}^i(\mu_{\Sigma})}{T^d_{pc}(0)}=1-\kappa\left(\frac{\mu_{\Sigma}}{T_{pc}^d(0)}\right)^2,
\end{eqnarray}

\noindent which is valid for low values of $\mu_{\Sigma}$. The index $i=d,c$ denotes the pseudocritical temperature of deconfinement and chiral phase transitions respectively. Different lattice parameters, such as temporal and spatial sizes, and different meson mass ratios $M_{ps}/M_v$ are considered. The extrapolated result for $M_{ps}/M_v=0.175$ indicates $\kappa\sim 0.06$ for both phase transitions. In the case of EPNJL model, we find $\kappa\sim 0.03$ for both transitions, which is significantly different from the LQCD result. We believe this discrepancy may be due to the lattice approach, which uses non-renormalized susceptibilities of the chiral condensate and Polyakov loop. This method may shift the peak positions compared to the renormalized case.

\begin{figure}[htb]
\vspace{0.75cm}
\includegraphics[scale=0.44]{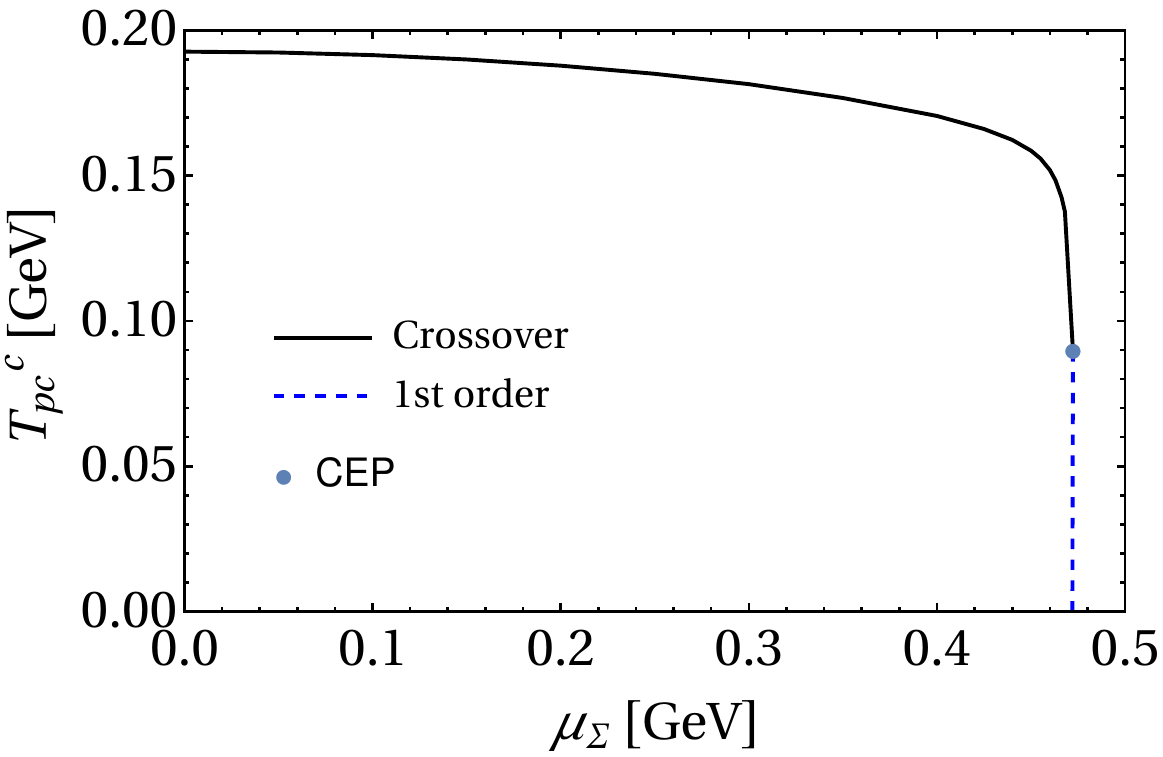}
\caption{Pseudocritical temperature as a function of the spin potential, $T_{pc}^c\times\mu_{\Sigma}$, for the chiral symmetry restoration. The crossover and first-order phase transition concerns to the continuous and dashed lines, respectively. These two regions are divided by the CEP, in the blue dot.}
\label{phase_diagram}
\end{figure}

In Fig. \ref{phase_diagram} we show the phase diagram (the pseudocritical temperature as a function of the spin potential) for the chiral symmetry restoration $T_{pc}^c \times \mu_{\Sigma}$. Then, around $T\sim 0.150$ GeV, a sharp drop leads to the critical end point (CEP) $(T_{CEP},\mu_{{\Sigma}_{CEP}})=(0.091\,\text{GeV},0.47226 \,\text{GeV})$. The first-order region shows a
steep descent, indicating that the spin potential changes slowly in this regime.

 Since the LQCD results in Ref.~\cite{Braguta:2025ddq} are limited to small $\mu_{\Sigma}$, this study provides, to the best of our knowledge, the first prediction of the phase diagram in the presence of a spin potential.

%%%%%%%%%%%%%%%%%%%%%%%%%%%%%%%%%%%%%%%%%%%
%%%%%%%%%%%%%%%%%%%%%%%%%%%%%%%%%

\section{Conclusions}
\label{conclusions}

In this work we explored the effects of the spin potential in the entangled--Polyakov--Nambu--Jona-Lasinio SU(2) in the mean-field approximation. The finite spin density is introduced
by the quark spin potential in the canonical formulation of the spin operator. We start by solving the partition function of the two-flavor Nambu--Jona-Lasinio model with the inclusion of the spin potential. Then, we obtain the quark dispersion relation, which is applied to the EPNJL model. To the best of our knowledge these is the first application of quark spin potential in a effective model of QCD, and our approach can be applied to different models.

In the numerical results we observe a decrease of the effective quark masses as function of temperature and the spin potential, indicating that both effects combined can restore partially the chiral symmetry. At low $T=0$ and $\mu_{\Sigma}=0.472$ GeV we observe a sharp decrease of the quark masses indicating a first-order phase transition. The expectation value of the Polyakov loop has been analyzed indicating that the deconfinement transition is shifted to lower values of temperature as we increase the spin potential.

By looking to the peak of chiral and deconfinement susceptibilities we can find the pseudocritical temperatures. In the EPNJL model, both pseudocritical temperatures are very close, which is confirmed by looking the the phase diagram. When applying the quadratic fitting formula for $T_{pc}(\mu_{\Sigma})$, we obtain $\kappa\sim 0.03$ for both pseudocritical temperatures, in disagreement with the continuum extrapolated value of LQCD, i.e., $\kappa\sim0.06$. However the qualitative behavior is preserved where pseudocritical temperatures for chiral and deconfinement phase transitions decrease as we increase the spin potential. The disagreement related to the curvature of the phase diagram can be related to model limitations or due the lattice results used nonrenormalized susceptibilities of
the chiral condensate and Polyakov loop.

As a prediction, we show the phase diagram for the chiral symmetry restoration    $T^c_{pc}\times \mu_{\Sigma}$. Besides, lattice QCD can work with low values of $\mu_{\Sigma}$~\cite{Braguta:2025ddq}, after an analytical continuation from imaginary to real values of spin potential. The advantage of our prescription is to access all values of interest temperature and spin potential. The phase diagram indicates a region, at $T\lesssim 0.150$ GeV where the pseudocritical decreases smoothly, and then suffers a sharp decrease until the CEP. After this point, we observe a line of first-order phase transition temperatures which changes in a very slow way as a function of $\mu_{\Sigma}$.

The NJL model, along with its extensions, is most often applied within the mean-field approximation. While widely used, this approach can be problematic for determining critical exponents, as suggested by universality arguments. When the model is confronted with data from compact objects, for example, one finds that the expected values of the sound velocity~\cite{PhysRevD.105.023018,PhysRevC.107.045807} cannot be reproduced without further refinements. In our case, however, external constraints must be more clearly understood, given the still incipient nature of the recently proposed spin potential.
It is also well-established that going beyond the mean-field level modifies the transition temperature as a function of the quark chemical potential, $T \times \mu$~\cite{PhysRevC.81.065205}. Nevertheless, effective models remain powerful tools for exploring key features of quark matter under extreme conditions, as exemplified by the NJL model~\cite{Klevansky:1992qe,Buballa:2003qv}. In the seminal LQCD study introducing the spin potential~\cite{Braguta:2025ddq}, the analysis focuses solely on quark degrees of freedom, which supports the view that effective quark models are well-suited for studying such systems and making comparisons. An additional advantage of the present approach is that it allows us to make predictions in the high $\mu_{\Sigma}$ regime, since the sign problem is absent---at least in the range 
$\mu_{\Sigma} < \Lambda$. For the purposes of the present work, the mean-field approximation thus provides a coherent qualitative picture of the underlying physics, in agreement with LQCD trends. Future extensions beyond the mean-field level, as well as refinements of the coupling, will require more robust analytical and numerical developments, which are already part of our research plans. We are currently applying this approach to a renormalized theory and comparing the results with lattice QCD data. The renormalization procedure, based on the dispersion relation derived in this work, requires careful implementation. These findings, along with further analyses, will be reported in a forthcoming study.

\section*{Acknowledgments}

This work was partially supported by Conselho Nacional de Desenvolvimento Cient\'ifico e Tecno\-l\'o\-gico  (CNPq), Grants No. 312032/2023-4, 402963/2024-5 and 445182/2024-5 (R.L.S.F.); Fundação Carlos Chagas Filho de Amparo à Pesquisa do Estado do Rio de Janeiro (FAPERJ), Grant No.SEI-260003/019544/2022 (W.R.T);
Funda\c{c}\~ao de Amparo \`a Pesquisa do Estado do Rio 
Grande do Sul (FAPERGS), Grants Nos. 24/2551-0001285-0 and 23/2551-0000791-6 (R.L.S.F.); The work is also part of the project Instituto
Nacional de Ciência e Tecnologia—Física Nuclear e
Aplicações (INCT—FNA), Grant No. 464898/2014-5
and supported by the Serrapilheira Institute (Grant
No. Serra-2211-42230).

\bibliography{ref.bib}

\begin{thebibliography}{75}%
\makeatletter
\providecommand \@ifxundefined [1]{%
 \@ifx{#1\undefined}
}%
\providecommand \@ifnum [1]{%
 \ifnum #1\expandafter \@firstoftwo
 \else \expandafter \@secondoftwo
 \fi
}%
\providecommand \@ifx [1]{%
 \ifx #1\expandafter \@firstoftwo
 \else \expandafter \@secondoftwo
 \fi
}%
\providecommand \natexlab [1]{#1}%
\providecommand \enquote  [1]{``#1''}%
\providecommand \bibnamefont  [1]{#1}%
\providecommand \bibfnamefont [1]{#1}%
\providecommand \citenamefont [1]{#1}%
\providecommand \href@noop [0]{\@secondoftwo}%
\providecommand \href [0]{\begingroup \@sanitize@url \@href}%
\providecommand \@href[1]{\@@startlink{#1}\@@href}%
\providecommand \@@href[1]{\endgroup#1\@@endlink}%
\providecommand \@sanitize@url [0]{\catcode `\\12\catcode `\$12\catcode
  `\&12\catcode `\#12\catcode `\^12\catcode `\_12\catcode `\%12\relax}%
\providecommand \@@startlink[1]{}%
\providecommand \@@endlink[0]{}%
\providecommand \url  [0]{\begingroup\@sanitize@url \@url }%
\providecommand \@url [1]{\endgroup\@href {#1}{\urlprefix }}%
\providecommand \urlprefix  [0]{URL }%
\providecommand \Eprint [0]{\href }%
\providecommand \doibase [0]{http://dx.doi.org/}%
\providecommand \selectlanguage [0]{\@gobble}%
\providecommand \bibinfo  [0]{\@secondoftwo}%
\providecommand \bibfield  [0]{\@secondoftwo}%
\providecommand \translation [1]{[#1]}%
\providecommand \BibitemOpen [0]{}%
\providecommand \bibitemStop [0]{}%
\providecommand \bibitemNoStop [0]{.\EOS\space}%
\providecommand \EOS [0]{\spacefactor3000\relax}%
\providecommand \BibitemShut  [1]{\csname bibitem#1\endcsname}%
\let\auto@bib@innerbib\@empty
%</preamble>
\bibitem [{\citenamefont {Busza}\ \emph {et~al.}(2018)\citenamefont {Busza},
  \citenamefont {Rajagopal},\ and\ \citenamefont {van~der
  Schee}}]{Busza:2018rrf}%
  \BibitemOpen
  \bibfield  {author} {\bibinfo {author} {\bibfnamefont {W.}~\bibnamefont
  {Busza}}, \bibinfo {author} {\bibfnamefont {K.}~\bibnamefont {Rajagopal}}, \
  and\ \bibinfo {author} {\bibfnamefont {W.}~\bibnamefont {van~der Schee}},\
  }\href {\doibase 10.1146/annurev-nucl-101917-020852} {\bibfield  {journal}
  {\bibinfo  {journal} {Ann. Rev. Nucl. Part. Sci.}\ }\textbf {\bibinfo
  {volume} {68}},\ \bibinfo {pages} {339} (\bibinfo {year} {2018})},\ \Eprint
  {http://arxiv.org/abs/1802.04801} {arXiv:1802.04801 [hep-ph]} \BibitemShut
  {NoStop}%
\bibitem [{\citenamefont {Adamczyk}\ \emph {et~al.}(2017)\citenamefont
  {Adamczyk} \emph {et~al.}}]{STAR:2017ckg}%
  \BibitemOpen
  \bibfield  {author} {\bibinfo {author} {\bibfnamefont {L.}~\bibnamefont
  {Adamczyk}} \emph {et~al.} (\bibinfo {collaboration} {STAR}),\ }\href
  {\doibase 10.1038/nature23004} {\bibfield  {journal} {\bibinfo  {journal}
  {Nature}\ }\textbf {\bibinfo {volume} {548}},\ \bibinfo {pages} {62}
  (\bibinfo {year} {2017})},\ \Eprint {http://arxiv.org/abs/1701.06657}
  {arXiv:1701.06657 [nucl-ex]} \BibitemShut {NoStop}%
\bibitem [{\citenamefont {Jiang}\ \emph {et~al.}(2016)\citenamefont {Jiang},
  \citenamefont {Lin},\ and\ \citenamefont {Liao}}]{Jiang:2016woz}%
  \BibitemOpen
  \bibfield  {author} {\bibinfo {author} {\bibfnamefont {Y.}~\bibnamefont
  {Jiang}}, \bibinfo {author} {\bibfnamefont {Z.-W.}\ \bibnamefont {Lin}}, \
  and\ \bibinfo {author} {\bibfnamefont {J.}~\bibnamefont {Liao}},\ }\href
  {\doibase 10.1103/PhysRevC.94.044910} {\bibfield  {journal} {\bibinfo
  {journal} {Phys. Rev. C}\ }\textbf {\bibinfo {volume} {94}},\ \bibinfo
  {pages} {044910} (\bibinfo {year} {2016})},\ \bibinfo {note} {[Erratum:
  Phys.Rev.C 95, 049904 (2017)]},\ \Eprint {http://arxiv.org/abs/1602.06580}
  {arXiv:1602.06580 [hep-ph]} \BibitemShut {NoStop}%
\bibitem [{\citenamefont {Acharya}\ \emph {et~al.}(2020)\citenamefont {Acharya}
  \emph {et~al.}}]{ALICE:2019aid}%
  \BibitemOpen
  \bibfield  {author} {\bibinfo {author} {\bibfnamefont {S.}~\bibnamefont
  {Acharya}} \emph {et~al.} (\bibinfo {collaboration} {ALICE}),\ }\href
  {\doibase 10.1103/PhysRevLett.125.012301} {\bibfield  {journal} {\bibinfo
  {journal} {Phys. Rev. Lett.}\ }\textbf {\bibinfo {volume} {125}},\ \bibinfo
  {pages} {012301} (\bibinfo {year} {2020})},\ \Eprint
  {http://arxiv.org/abs/1910.14408} {arXiv:1910.14408 [nucl-ex]} \BibitemShut
  {NoStop}%
\bibitem [{\citenamefont {Abdallah}\ \emph {et~al.}(2023)\citenamefont
  {Abdallah} \emph {et~al.}}]{STAR:2022fan}%
  \BibitemOpen
  \bibfield  {author} {\bibinfo {author} {\bibfnamefont {M.~S.}\ \bibnamefont
  {Abdallah}} \emph {et~al.} (\bibinfo {collaboration} {STAR}),\ }\href
  {\doibase 10.1038/s41586-022-05557-5} {\bibfield  {journal} {\bibinfo
  {journal} {Nature}\ }\textbf {\bibinfo {volume} {614}},\ \bibinfo {pages}
  {244} (\bibinfo {year} {2023})},\ \Eprint {http://arxiv.org/abs/2204.02302}
  {arXiv:2204.02302 [hep-ph]} \BibitemShut {NoStop}%
\bibitem [{\citenamefont {Braguta}\ \emph {et~al.}(2022)\citenamefont
  {Braguta}, \citenamefont {Kotov}, \citenamefont {Kuznedelev},\ and\
  \citenamefont {Roenko}}]{Braguta:2021ucr}%
  \BibitemOpen
  \bibfield  {author} {\bibinfo {author} {\bibfnamefont {V.}~\bibnamefont
  {Braguta}}, \bibinfo {author} {\bibfnamefont {A.~Y.}\ \bibnamefont {Kotov}},
  \bibinfo {author} {\bibfnamefont {D.}~\bibnamefont {Kuznedelev}}, \ and\
  \bibinfo {author} {\bibfnamefont {A.}~\bibnamefont {Roenko}},\ }\href
  {\doibase 10.22323/1.396.0125} {\bibfield  {journal} {\bibinfo  {journal}
  {PoS}\ }\textbf {\bibinfo {volume} {LATTICE2021}},\ \bibinfo {pages} {125}
  (\bibinfo {year} {2022})},\ \Eprint {http://arxiv.org/abs/2110.12302}
  {arXiv:2110.12302 [hep-lat]} \BibitemShut {NoStop}%
\bibitem [{\citenamefont {Braguta}\ \emph {et~al.}(2021)\citenamefont
  {Braguta}, \citenamefont {Kotov}, \citenamefont {Kuznedelev},\ and\
  \citenamefont {Roenko}}]{Braguta:2021jgn}%
  \BibitemOpen
  \bibfield  {author} {\bibinfo {author} {\bibfnamefont {V.~V.}\ \bibnamefont
  {Braguta}}, \bibinfo {author} {\bibfnamefont {A.~Y.}\ \bibnamefont {Kotov}},
  \bibinfo {author} {\bibfnamefont {D.~D.}\ \bibnamefont {Kuznedelev}}, \ and\
  \bibinfo {author} {\bibfnamefont {A.~A.}\ \bibnamefont {Roenko}},\ }\href
  {\doibase 10.1103/PhysRevD.103.094515} {\bibfield  {journal} {\bibinfo
  {journal} {Phys. Rev. D}\ }\textbf {\bibinfo {volume} {103}},\ \bibinfo
  {pages} {094515} (\bibinfo {year} {2021})},\ \Eprint
  {http://arxiv.org/abs/2102.05084} {arXiv:2102.05084 [hep-lat]} \BibitemShut
  {NoStop}%
\bibitem [{\citenamefont {Braguta}\ \emph
  {et~al.}(2023{\natexlab{a}})\citenamefont {Braguta}, \citenamefont {Kotov},
  \citenamefont {Roenko},\ and\ \citenamefont {Sychev}}]{Braguta:2022str}%
  \BibitemOpen
  \bibfield  {author} {\bibinfo {author} {\bibfnamefont {V.~V.}\ \bibnamefont
  {Braguta}}, \bibinfo {author} {\bibfnamefont {A.}~\bibnamefont {Kotov}},
  \bibinfo {author} {\bibfnamefont {A.}~\bibnamefont {Roenko}}, \ and\ \bibinfo
  {author} {\bibfnamefont {D.}~\bibnamefont {Sychev}},\ }\href {\doibase
  10.22323/1.430.0190} {\bibfield  {journal} {\bibinfo  {journal} {PoS}\
  }\textbf {\bibinfo {volume} {LATTICE2022}},\ \bibinfo {pages} {190} (\bibinfo
  {year} {2023}{\natexlab{a}})},\ \Eprint {http://arxiv.org/abs/2212.03224}
  {arXiv:2212.03224 [hep-lat]} \BibitemShut {NoStop}%
\bibitem [{\citenamefont {Yang}\ and\ \citenamefont
  {Huang}(2023)}]{Yang:2023vsw}%
  \BibitemOpen
  \bibfield  {author} {\bibinfo {author} {\bibfnamefont {J.-C.}\ \bibnamefont
  {Yang}}\ and\ \bibinfo {author} {\bibfnamefont {X.-G.}\ \bibnamefont
  {Huang}},\ }\href@noop {} {\  (\bibinfo {year} {2023})},\ \Eprint
  {http://arxiv.org/abs/arXiv: 2307.05755} {arXiv:arXiv: 2307.05755 [hep-lat]}
  \BibitemShut {NoStop}%
\bibitem [{\citenamefont {Braguta}\ \emph
  {et~al.}(2023{\natexlab{b}})\citenamefont {Braguta}, \citenamefont
  {Chernodub}, \citenamefont {Kudrov}, \citenamefont {Roenko},\ and\
  \citenamefont {Sychev}}]{Braguta:2023aio}%
  \BibitemOpen
  \bibfield  {author} {\bibinfo {author} {\bibfnamefont {V.~V.}\ \bibnamefont
  {Braguta}}, \bibinfo {author} {\bibfnamefont {M.~N.}\ \bibnamefont
  {Chernodub}}, \bibinfo {author} {\bibfnamefont {I.~E.}\ \bibnamefont
  {Kudrov}}, \bibinfo {author} {\bibfnamefont {A.~A.}\ \bibnamefont {Roenko}},
  \ and\ \bibinfo {author} {\bibfnamefont {D.~A.}\ \bibnamefont {Sychev}},\
  }\href {\doibase 10.1134/S1063778824010150} {\bibfield  {journal} {\bibinfo
  {journal} {Phys. Atom. Nucl.}\ }\textbf {\bibinfo {volume} {86}},\ \bibinfo
  {pages} {1249} (\bibinfo {year} {2023}{\natexlab{b}})}\BibitemShut {NoStop}%
\bibitem [{\citenamefont {Chernodub}\ \emph {et~al.}(2023)\citenamefont
  {Chernodub}, \citenamefont {Goy},\ and\ \citenamefont
  {Molochkov}}]{Chernodub:2022veq}%
  \BibitemOpen
  \bibfield  {author} {\bibinfo {author} {\bibfnamefont {M.~N.}\ \bibnamefont
  {Chernodub}}, \bibinfo {author} {\bibfnamefont {V.~A.}\ \bibnamefont {Goy}},
  \ and\ \bibinfo {author} {\bibfnamefont {A.~V.}\ \bibnamefont {Molochkov}},\
  }\href {\doibase 10.1103/PhysRevD.107.114502} {\bibfield  {journal} {\bibinfo
   {journal} {Phys. Rev. D}\ }\textbf {\bibinfo {volume} {107}},\ \bibinfo
  {pages} {114502} (\bibinfo {year} {2023})},\ \Eprint
  {http://arxiv.org/abs/2209.15534} {arXiv:2209.15534 [hep-lat]} \BibitemShut
  {NoStop}%
\bibitem [{\citenamefont {Jiang}\ and\ \citenamefont
  {Liao}(2016)}]{Jiang:2016wvv}%
  \BibitemOpen
  \bibfield  {author} {\bibinfo {author} {\bibfnamefont {Y.}~\bibnamefont
  {Jiang}}\ and\ \bibinfo {author} {\bibfnamefont {J.}~\bibnamefont {Liao}},\
  }\href {\doibase 10.1103/PhysRevLett.117.192302} {\bibfield  {journal}
  {\bibinfo  {journal} {Phys. Rev. Lett.}\ }\textbf {\bibinfo {volume} {117}},\
  \bibinfo {pages} {192302} (\bibinfo {year} {2016})},\ \Eprint
  {http://arxiv.org/abs/1606.03808} {arXiv:1606.03808 [hep-ph]} \BibitemShut
  {NoStop}%
\bibitem [{\citenamefont {Chernodub}\ and\ \citenamefont
  {Gongyo}(2017)}]{Chernodub:2016kxh}%
  \BibitemOpen
  \bibfield  {author} {\bibinfo {author} {\bibfnamefont {M.~N.}\ \bibnamefont
  {Chernodub}}\ and\ \bibinfo {author} {\bibfnamefont {S.}~\bibnamefont
  {Gongyo}},\ }\href {\doibase 10.1007/JHEP01(2017)136} {\bibfield  {journal}
  {\bibinfo  {journal} {JHEP}\ }\textbf {\bibinfo {volume} {01}},\ \bibinfo
  {pages} {136} (\bibinfo {year} {2017})},\ \Eprint
  {http://arxiv.org/abs/1611.02598} {arXiv:1611.02598 [hep-th]} \BibitemShut
  {NoStop}%
\bibitem [{\citenamefont {Sun}\ and\ \citenamefont
  {Huang}(2022)}]{Sun:2021hxo}%
  \BibitemOpen
  \bibfield  {author} {\bibinfo {author} {\bibfnamefont {F.}~\bibnamefont
  {Sun}}\ and\ \bibinfo {author} {\bibfnamefont {A.}~\bibnamefont {Huang}},\
  }\href {\doibase 10.1103/PhysRevD.106.076007} {\bibfield  {journal} {\bibinfo
   {journal} {Phys. Rev. D}\ }\textbf {\bibinfo {volume} {106}},\ \bibinfo
  {pages} {076007} (\bibinfo {year} {2022})},\ \Eprint
  {http://arxiv.org/abs/2104.14382} {arXiv:2104.14382 [hep-ph]} \BibitemShut
  {NoStop}%
\bibitem [{\citenamefont {Sun}\ \emph {et~al.}(2023{\natexlab{a}})\citenamefont
  {Sun}, \citenamefont {Li}, \citenamefont {Wen}, \citenamefont {Huang},\ and\
  \citenamefont {Xie}}]{Sun:2023yux}%
  \BibitemOpen
  \bibfield  {author} {\bibinfo {author} {\bibfnamefont {F.}~\bibnamefont
  {Sun}}, \bibinfo {author} {\bibfnamefont {S.}~\bibnamefont {Li}}, \bibinfo
  {author} {\bibfnamefont {R.}~\bibnamefont {Wen}}, \bibinfo {author}
  {\bibfnamefont {A.}~\bibnamefont {Huang}}, \ and\ \bibinfo {author}
  {\bibfnamefont {W.}~\bibnamefont {Xie}},\ }\href@noop {} {\  (\bibinfo {year}
  {2023}{\natexlab{a}})},\ \Eprint {http://arxiv.org/abs/2310.18942}
  {arXiv:2310.18942 [hep-ph]} \BibitemShut {NoStop}%
\bibitem [{\citenamefont {Wang}\ \emph {et~al.}(2019)\citenamefont {Wang},
  \citenamefont {Wei}, \citenamefont {Li},\ and\ \citenamefont
  {Huang}}]{Wang:2018sur}%
  \BibitemOpen
  \bibfield  {author} {\bibinfo {author} {\bibfnamefont {X.}~\bibnamefont
  {Wang}}, \bibinfo {author} {\bibfnamefont {M.}~\bibnamefont {Wei}}, \bibinfo
  {author} {\bibfnamefont {Z.}~\bibnamefont {Li}}, \ and\ \bibinfo {author}
  {\bibfnamefont {M.}~\bibnamefont {Huang}},\ }\href {\doibase
  10.1103/PhysRevD.99.016018} {\bibfield  {journal} {\bibinfo  {journal} {Phys.
  Rev. D}\ }\textbf {\bibinfo {volume} {99}},\ \bibinfo {pages} {016018}
  (\bibinfo {year} {2019})},\ \Eprint {http://arxiv.org/abs/1808.01931}
  {arXiv:1808.01931 [hep-ph]} \BibitemShut {NoStop}%
\bibitem [{\citenamefont {Singha}\ \emph {et~al.}(2024)\citenamefont {Singha},
  \citenamefont {Ambrus},\ and\ \citenamefont {Chernodub}}]{Singha:2024tpo}%
  \BibitemOpen
  \bibfield  {author} {\bibinfo {author} {\bibfnamefont {P.}~\bibnamefont
  {Singha}}, \bibinfo {author} {\bibfnamefont {V.~E.}\ \bibnamefont {Ambrus}},
  \ and\ \bibinfo {author} {\bibfnamefont {M.~N.}\ \bibnamefont {Chernodub}},\
  }\href {\doibase 10.1103/PhysRevD.110.094053} {\bibfield  {journal} {\bibinfo
   {journal} {Phys. Rev. D}\ }\textbf {\bibinfo {volume} {110}},\ \bibinfo
  {pages} {094053} (\bibinfo {year} {2024})},\ \Eprint
  {http://arxiv.org/abs/2407.07828} {arXiv:2407.07828 [hep-ph]} \BibitemShut
  {NoStop}%
\bibitem [{\citenamefont {Hern\'andez}\ and\ \citenamefont
  {Zamora}(2024)}]{Hernandez:2024nev}%
  \BibitemOpen
  \bibfield  {author} {\bibinfo {author} {\bibfnamefont {L.~A.}\ \bibnamefont
  {Hern\'andez}}\ and\ \bibinfo {author} {\bibfnamefont {R.}~\bibnamefont
  {Zamora}},\ }\href@noop {} {\  (\bibinfo {year} {2024})},\ \Eprint
  {http://arxiv.org/abs/arXiv: 2410.17874} {arXiv:arXiv: 2410.17874 [hep-ph]}
  \BibitemShut {NoStop}%
\bibitem [{\citenamefont {Braguta}\ \emph
  {et~al.}(2024{\natexlab{a}})\citenamefont {Braguta}, \citenamefont
  {Chernodub}, \citenamefont {Kudrov}, \citenamefont {Roenko},\ and\
  \citenamefont {Sychev}}]{Braguta:2023tqz}%
  \BibitemOpen
  \bibfield  {author} {\bibinfo {author} {\bibfnamefont {V.~V.}\ \bibnamefont
  {Braguta}}, \bibinfo {author} {\bibfnamefont {M.~N.}\ \bibnamefont
  {Chernodub}}, \bibinfo {author} {\bibfnamefont {I.~E.}\ \bibnamefont
  {Kudrov}}, \bibinfo {author} {\bibfnamefont {A.~A.}\ \bibnamefont {Roenko}},
  \ and\ \bibinfo {author} {\bibfnamefont {D.~A.}\ \bibnamefont {Sychev}},\
  }\href {\doibase 10.1103/PhysRevD.110.014511} {\bibfield  {journal} {\bibinfo
   {journal} {Phys. Rev. D}\ }\textbf {\bibinfo {volume} {110}},\ \bibinfo
  {pages} {014511} (\bibinfo {year} {2024}{\natexlab{a}})},\ \Eprint
  {http://arxiv.org/abs/2310.16036} {arXiv:2310.16036 [hep-ph]} \BibitemShut
  {NoStop}%
\bibitem [{\citenamefont {Braguta}\ \emph
  {et~al.}(2024{\natexlab{b}})\citenamefont {Braguta}, \citenamefont
  {Chernodub}, \citenamefont {Roenko},\ and\ \citenamefont
  {Sychev}}]{Braguta:2023yjn}%
  \BibitemOpen
  \bibfield  {author} {\bibinfo {author} {\bibfnamefont {V.~V.}\ \bibnamefont
  {Braguta}}, \bibinfo {author} {\bibfnamefont {M.~N.}\ \bibnamefont
  {Chernodub}}, \bibinfo {author} {\bibfnamefont {A.~A.}\ \bibnamefont
  {Roenko}}, \ and\ \bibinfo {author} {\bibfnamefont {D.~A.}\ \bibnamefont
  {Sychev}},\ }\href {\doibase 10.1016/j.physletb.2024.138604} {\bibfield
  {journal} {\bibinfo  {journal} {Phys. Lett. B}\ }\textbf {\bibinfo {volume}
  {852}},\ \bibinfo {pages} {138604} (\bibinfo {year} {2024}{\natexlab{b}})},\
  \Eprint {http://arxiv.org/abs/2303.03147} {arXiv:2303.03147 [hep-lat]}
  \BibitemShut {NoStop}%
\bibitem [{\citenamefont {Braguta}\ \emph
  {et~al.}(2024{\natexlab{c}})\citenamefont {Braguta}, \citenamefont
  {Chernodub}, \citenamefont {Kudrov}, \citenamefont {Roenko},\ and\
  \citenamefont {Sychev}}]{Braguta:2023qex}%
  \BibitemOpen
  \bibfield  {author} {\bibinfo {author} {\bibfnamefont {V.~V.}\ \bibnamefont
  {Braguta}}, \bibinfo {author} {\bibfnamefont {M.~N.}\ \bibnamefont
  {Chernodub}}, \bibinfo {author} {\bibfnamefont {I.~E.}\ \bibnamefont
  {Kudrov}}, \bibinfo {author} {\bibfnamefont {A.~A.}\ \bibnamefont {Roenko}},
  \ and\ \bibinfo {author} {\bibfnamefont {D.~A.}\ \bibnamefont {Sychev}},\
  }\href {\doibase 10.22323/1.453.0181} {\bibfield  {journal} {\bibinfo
  {journal} {PoS}\ }\textbf {\bibinfo {volume} {LATTICE2023}},\ \bibinfo
  {pages} {181} (\bibinfo {year} {2024}{\natexlab{c}})},\ \Eprint
  {http://arxiv.org/abs/2311.03947} {arXiv:2311.03947 [hep-lat]} \BibitemShut
  {NoStop}%
\bibitem [{\citenamefont {Jiang}(2022)}]{Jiang:2021izj}%
  \BibitemOpen
  \bibfield  {author} {\bibinfo {author} {\bibfnamefont {Y.}~\bibnamefont
  {Jiang}},\ }\href {\doibase 10.1140/epjc/s10052-022-10915-8} {\bibfield
  {journal} {\bibinfo  {journal} {Eur. Phys. J. C}\ }\textbf {\bibinfo {volume}
  {82}},\ \bibinfo {pages} {949} (\bibinfo {year} {2022})},\ \Eprint
  {http://arxiv.org/abs/2108.09622} {arXiv:2108.09622 [hep-ph]} \BibitemShut
  {NoStop}%
\bibitem [{\citenamefont {Nunes}\ \emph {et~al.}(2025)\citenamefont {Nunes},
  \citenamefont {Farias}, \citenamefont {Tavares},\ and\ \citenamefont
  {Tim{\'o}teo}}]{Nunes:2024hzy}%
  \BibitemOpen
  \bibfield  {author} {\bibinfo {author} {\bibfnamefont {R.~M.}\ \bibnamefont
  {Nunes}}, \bibinfo {author} {\bibfnamefont {R.~L.~S.}\ \bibnamefont
  {Farias}}, \bibinfo {author} {\bibfnamefont {W.~R.}\ \bibnamefont {Tavares}},
  \ and\ \bibinfo {author} {\bibfnamefont {V.~S.}\ \bibnamefont
  {Tim{\'o}teo}},\ }\href {\doibase 10.1103/PhysRevD.111.056026} {\bibfield
  {journal} {\bibinfo  {journal} {Phys. Rev. D}\ }\textbf {\bibinfo {volume}
  {111}},\ \bibinfo {pages} {056026} (\bibinfo {year} {2025})},\ \Eprint
  {http://arxiv.org/abs/2412.14541} {arXiv:2412.14541 [hep-ph]} \BibitemShut
  {NoStop}%
\bibitem [{\citenamefont {Chen}\ \emph
  {et~al.}(2021{\natexlab{a}})\citenamefont {Chen}, \citenamefont {Huang},\
  and\ \citenamefont {Liao}}]{Chen:2021aiq}%
  \BibitemOpen
  \bibfield  {author} {\bibinfo {author} {\bibfnamefont {H.-L.}\ \bibnamefont
  {Chen}}, \bibinfo {author} {\bibfnamefont {X.-G.}\ \bibnamefont {Huang}}, \
  and\ \bibinfo {author} {\bibfnamefont {J.}~\bibnamefont {Liao}},\ }\href
  {\doibase 10.1007/978-3-030-71427-7_11} {\bibfield  {journal} {\bibinfo
  {journal} {Lect. Notes Phys.}\ }\textbf {\bibinfo {volume} {987}},\ \bibinfo
  {pages} {349} (\bibinfo {year} {2021}{\natexlab{a}})},\ \Eprint
  {http://arxiv.org/abs/2108.00586} {arXiv:2108.00586 [hep-ph]} \BibitemShut
  {NoStop}%
\bibitem [{\citenamefont {Chernodub}(2021)}]{Chernodub:2020qah}%
  \BibitemOpen
  \bibfield  {author} {\bibinfo {author} {\bibfnamefont {M.~N.}\ \bibnamefont
  {Chernodub}},\ }\href {\doibase 10.1103/PhysRevD.103.054027} {\bibfield
  {journal} {\bibinfo  {journal} {Phys. Rev. D}\ }\textbf {\bibinfo {volume}
  {103}},\ \bibinfo {pages} {054027} (\bibinfo {year} {2021})},\ \Eprint
  {http://arxiv.org/abs/2012.04924} {arXiv:2012.04924 [hep-ph]} \BibitemShut
  {NoStop}%
\bibitem [{\citenamefont {Braguta}\ \emph
  {et~al.}(2024{\natexlab{d}})\citenamefont {Braguta}, \citenamefont
  {Chernodub},\ and\ \citenamefont {Roenko}}]{Braguta:2023iyx}%
  \BibitemOpen
  \bibfield  {author} {\bibinfo {author} {\bibfnamefont {V.~V.}\ \bibnamefont
  {Braguta}}, \bibinfo {author} {\bibfnamefont {M.~N.}\ \bibnamefont
  {Chernodub}}, \ and\ \bibinfo {author} {\bibfnamefont {A.~A.}\ \bibnamefont
  {Roenko}},\ }\href {\doibase 10.1016/j.physletb.2024.138783} {\bibfield
  {journal} {\bibinfo  {journal} {Phys. Lett. B}\ }\textbf {\bibinfo {volume}
  {855}},\ \bibinfo {pages} {138783} (\bibinfo {year} {2024}{\natexlab{d}})},\
  \Eprint {http://arxiv.org/abs/2312.13994} {arXiv:2312.13994 [hep-lat]}
  \BibitemShut {NoStop}%
\bibitem [{\citenamefont {Sun}\ \emph {et~al.}(2024)\citenamefont {Sun},
  \citenamefont {Shao}, \citenamefont {Wen}, \citenamefont {Xu},\ and\
  \citenamefont {Huang}}]{Sun:2024anu}%
  \BibitemOpen
  \bibfield  {author} {\bibinfo {author} {\bibfnamefont {F.}~\bibnamefont
  {Sun}}, \bibinfo {author} {\bibfnamefont {J.}~\bibnamefont {Shao}}, \bibinfo
  {author} {\bibfnamefont {R.}~\bibnamefont {Wen}}, \bibinfo {author}
  {\bibfnamefont {K.}~\bibnamefont {Xu}}, \ and\ \bibinfo {author}
  {\bibfnamefont {M.}~\bibnamefont {Huang}},\ }\href {\doibase
  10.1103/PhysRevD.109.116017} {\bibfield  {journal} {\bibinfo  {journal}
  {Phys. Rev. D}\ }\textbf {\bibinfo {volume} {109}},\ \bibinfo {pages}
  {116017} (\bibinfo {year} {2024})},\ \Eprint
  {http://arxiv.org/abs/2402.16595} {arXiv:2402.16595 [hep-ph]} \BibitemShut
  {NoStop}%
\bibitem [{\citenamefont {Ghalati}\ and\ \citenamefont
  {Sadooghi}(2023)}]{Ghalati:2023npr}%
  \BibitemOpen
  \bibfield  {author} {\bibinfo {author} {\bibfnamefont {H.~M.}\ \bibnamefont
  {Ghalati}}\ and\ \bibinfo {author} {\bibfnamefont {N.}~\bibnamefont
  {Sadooghi}},\ }\href {\doibase 10.1103/PhysRevD.108.054032} {\bibfield
  {journal} {\bibinfo  {journal} {Phys. Rev. D}\ }\textbf {\bibinfo {volume}
  {108}},\ \bibinfo {pages} {054032} (\bibinfo {year} {2023})},\ \Eprint
  {http://arxiv.org/abs/2306.04472} {arXiv:2306.04472 [nucl-th]} \BibitemShut
  {NoStop}%
\bibitem [{\citenamefont {Sadooghi}\ \emph {et~al.}(2021)\citenamefont
  {Sadooghi}, \citenamefont {Tabatabaee~Mehr},\ and\ \citenamefont
  {Taghinavaz}}]{Sadooghi:2021upd}%
  \BibitemOpen
  \bibfield  {author} {\bibinfo {author} {\bibfnamefont {N.}~\bibnamefont
  {Sadooghi}}, \bibinfo {author} {\bibfnamefont {S.~M.~A.}\ \bibnamefont
  {Tabatabaee~Mehr}}, \ and\ \bibinfo {author} {\bibfnamefont {F.}~\bibnamefont
  {Taghinavaz}},\ }\href {\doibase 10.1103/PhysRevD.104.116022} {\bibfield
  {journal} {\bibinfo  {journal} {Phys. Rev. D}\ }\textbf {\bibinfo {volume}
  {104}},\ \bibinfo {pages} {116022} (\bibinfo {year} {2021})},\ \Eprint
  {http://arxiv.org/abs/2108.12760} {arXiv:2108.12760 [hep-ph]} \BibitemShut
  {NoStop}%
\bibitem [{\citenamefont {Hua}\ and\ \citenamefont {Feng}(2024)}]{Hua:2024bwn}%
  \BibitemOpen
  \bibfield  {author} {\bibinfo {author} {\bibfnamefont {Y.}~\bibnamefont
  {Hua}}\ and\ \bibinfo {author} {\bibfnamefont {S.-Q.}\ \bibnamefont {Feng}},\
  }\href@noop {} {\  (\bibinfo {year} {2024})},\ \Eprint
  {http://arxiv.org/abs/arXiv: 2412.06398} {arXiv:arXiv: 2412.06398 [hep-ph]}
  \BibitemShut {NoStop}%
\bibitem [{\citenamefont {Chen}\ \emph
  {et~al.}(2024{\natexlab{a}})\citenamefont {Chen}, \citenamefont {Fu},
  \citenamefont {Huang},\ and\ \citenamefont {Ma}}]{Chen:2024utf}%
  \BibitemOpen
  \bibfield  {author} {\bibinfo {author} {\bibfnamefont {H.-L.}\ \bibnamefont
  {Chen}}, \bibinfo {author} {\bibfnamefont {W.-j.}\ \bibnamefont {Fu}},
  \bibinfo {author} {\bibfnamefont {X.-G.}\ \bibnamefont {Huang}}, \ and\
  \bibinfo {author} {\bibfnamefont {G.-L.}\ \bibnamefont {Ma}},\ }\href@noop {}
  {\  (\bibinfo {year} {2024}{\natexlab{a}})},\ \Eprint
  {http://arxiv.org/abs/arXiv: 2410.20704} {arXiv:arXiv: 2410.20704 [hep-ph]}
  \BibitemShut {NoStop}%
\bibitem [{\citenamefont {Sun}\ \emph {et~al.}(2023{\natexlab{b}})\citenamefont
  {Sun}, \citenamefont {Xu},\ and\ \citenamefont {Huang}}]{Sun:2023kuu}%
  \BibitemOpen
  \bibfield  {author} {\bibinfo {author} {\bibfnamefont {F.}~\bibnamefont
  {Sun}}, \bibinfo {author} {\bibfnamefont {K.}~\bibnamefont {Xu}}, \ and\
  \bibinfo {author} {\bibfnamefont {M.}~\bibnamefont {Huang}},\ }\href
  {\doibase 10.1103/PhysRevD.108.096007} {\bibfield  {journal} {\bibinfo
  {journal} {Phys. Rev. D}\ }\textbf {\bibinfo {volume} {108}},\ \bibinfo
  {pages} {096007} (\bibinfo {year} {2023}{\natexlab{b}})},\ \Eprint
  {http://arxiv.org/abs/2307.14402} {arXiv:2307.14402 [hep-ph]} \BibitemShut
  {NoStop}%
\bibitem [{\citenamefont {Tabatabaee~Mehr}(2023)}]{TabatabaeeMehr:2023tpt}%
  \BibitemOpen
  \bibfield  {author} {\bibinfo {author} {\bibfnamefont {S.~M.~A.}\
  \bibnamefont {Tabatabaee~Mehr}},\ }\href {\doibase
  10.1103/PhysRevD.108.094042} {\bibfield  {journal} {\bibinfo  {journal}
  {Phys. Rev. D}\ }\textbf {\bibinfo {volume} {108}},\ \bibinfo {pages}
  {094042} (\bibinfo {year} {2023})},\ \Eprint
  {http://arxiv.org/abs/2306.11753} {arXiv:2306.11753 [nucl-th]} \BibitemShut
  {NoStop}%
\bibitem [{\citenamefont {Wei}\ and\ \citenamefont
  {Huang}(2023)}]{Wei:2023pdf}%
  \BibitemOpen
  \bibfield  {author} {\bibinfo {author} {\bibfnamefont {M.}~\bibnamefont
  {Wei}}\ and\ \bibinfo {author} {\bibfnamefont {M.}~\bibnamefont {Huang}},\
  }\href {\doibase 10.1088/1674-1137/acf036} {\bibfield  {journal} {\bibinfo
  {journal} {Chin. Phys. C}\ }\textbf {\bibinfo {volume} {47}},\ \bibinfo
  {pages} {104105} (\bibinfo {year} {2023})},\ \Eprint
  {http://arxiv.org/abs/2303.01897} {arXiv:2303.01897 [hep-ph]} \BibitemShut
  {NoStop}%
\bibitem [{\citenamefont {Mehr}\ and\ \citenamefont
  {Taghinavaz}(2023)}]{Mehr:2022tfq}%
  \BibitemOpen
  \bibfield  {author} {\bibinfo {author} {\bibfnamefont {S.~M. A.~T.}\
  \bibnamefont {Mehr}}\ and\ \bibinfo {author} {\bibfnamefont {F.}~\bibnamefont
  {Taghinavaz}},\ }\href {\doibase 10.1016/j.aop.2023.169357} {\bibfield
  {journal} {\bibinfo  {journal} {Annals Phys.}\ }\textbf {\bibinfo {volume}
  {454}},\ \bibinfo {pages} {169357} (\bibinfo {year} {2023})},\ \Eprint
  {http://arxiv.org/abs/2201.05398} {arXiv:2201.05398 [hep-ph]} \BibitemShut
  {NoStop}%
\bibitem [{\citenamefont {Wei}\ \emph {et~al.}(2022)\citenamefont {Wei},
  \citenamefont {Jiang},\ and\ \citenamefont {Huang}}]{wei2022mass}%
  \BibitemOpen
  \bibfield  {author} {\bibinfo {author} {\bibfnamefont {M.}~\bibnamefont
  {Wei}}, \bibinfo {author} {\bibfnamefont {Y.}~\bibnamefont {Jiang}}, \ and\
  \bibinfo {author} {\bibfnamefont {M.}~\bibnamefont {Huang}},\ }\href@noop {}
  {\bibfield  {journal} {\bibinfo  {journal} {Chinese Physics C}\ }\textbf
  {\bibinfo {volume} {46}},\ \bibinfo {pages} {024102} (\bibinfo {year}
  {2022})}\BibitemShut {NoStop}%
\bibitem [{\citenamefont {Chen}\ \emph
  {et~al.}(2024{\natexlab{b}})\citenamefont {Chen}, \citenamefont {Huang},\
  and\ \citenamefont {Mameda}}]{Chen:2019tcp}%
  \BibitemOpen
  \bibfield  {author} {\bibinfo {author} {\bibfnamefont {H.-L.}\ \bibnamefont
  {Chen}}, \bibinfo {author} {\bibfnamefont {X.-G.}\ \bibnamefont {Huang}}, \
  and\ \bibinfo {author} {\bibfnamefont {K.}~\bibnamefont {Mameda}},\ }\href
  {\doibase 10.1007/JHEP02(2024)216} {\bibfield  {journal} {\bibinfo  {journal}
  {JHEP}\ }\textbf {\bibinfo {volume} {02}},\ \bibinfo {pages} {216} (\bibinfo
  {year} {2024}{\natexlab{b}})},\ \Eprint {http://arxiv.org/abs/1910.02700}
  {arXiv:1910.02700 [nucl-th]} \BibitemShut {NoStop}%
\bibitem [{\citenamefont {Zhang}\ \emph {et~al.}(2020)\citenamefont {Zhang},
  \citenamefont {Hou},\ and\ \citenamefont {Liao}}]{Zhang:2018ome}%
  \BibitemOpen
  \bibfield  {author} {\bibinfo {author} {\bibfnamefont {H.}~\bibnamefont
  {Zhang}}, \bibinfo {author} {\bibfnamefont {D.}~\bibnamefont {Hou}}, \ and\
  \bibinfo {author} {\bibfnamefont {J.}~\bibnamefont {Liao}},\ }\href {\doibase
  10.1088/1674-1137/abae4d} {\bibfield  {journal} {\bibinfo  {journal} {Chin.
  Phys. C}\ }\textbf {\bibinfo {volume} {44}},\ \bibinfo {pages} {111001}
  (\bibinfo {year} {2020})},\ \Eprint {http://arxiv.org/abs/1812.11787}
  {arXiv:1812.11787 [hep-ph]} \BibitemShut {NoStop}%
\bibitem [{\citenamefont {Cao}(2024)}]{Cao:2023olg}%
  \BibitemOpen
  \bibfield  {author} {\bibinfo {author} {\bibfnamefont {G.}~\bibnamefont
  {Cao}},\ }\href {\doibase 10.1103/PhysRevD.109.014001} {\bibfield  {journal}
  {\bibinfo  {journal} {Phys. Rev. D}\ }\textbf {\bibinfo {volume} {109}},\
  \bibinfo {pages} {014001} (\bibinfo {year} {2024})},\ \Eprint
  {http://arxiv.org/abs/2310.03310} {arXiv:2310.03310 [nucl-th]} \BibitemShut
  {NoStop}%
\bibitem [{\citenamefont {Chernodub}\ \emph {et~al.}(2024)\citenamefont
  {Chernodub}, \citenamefont {Goy}, \citenamefont {Molochkov}, \citenamefont
  {Stepanov},\ and\ \citenamefont {Pochinok}}]{Chernodub:2024wis}%
  \BibitemOpen
  \bibfield  {author} {\bibinfo {author} {\bibfnamefont {M.~N.}\ \bibnamefont
  {Chernodub}}, \bibinfo {author} {\bibfnamefont {V.~A.}\ \bibnamefont {Goy}},
  \bibinfo {author} {\bibfnamefont {A.~V.}\ \bibnamefont {Molochkov}}, \bibinfo
  {author} {\bibfnamefont {D.~V.}\ \bibnamefont {Stepanov}}, \ and\ \bibinfo
  {author} {\bibfnamefont {A.~S.}\ \bibnamefont {Pochinok}},\ }\href@noop {} {\
   (\bibinfo {year} {2024})},\ \Eprint {http://arxiv.org/abs/2409.01847}
  {arXiv:2409.01847 [hep-lat]} \BibitemShut {NoStop}%
\bibitem [{\citenamefont {Chen}\ \emph {et~al.}(2023)\citenamefont {Chen},
  \citenamefont {Zhu},\ and\ \citenamefont {Huang}}]{Chen:2023cjt}%
  \BibitemOpen
  \bibfield  {author} {\bibinfo {author} {\bibfnamefont {H.-L.}\ \bibnamefont
  {Chen}}, \bibinfo {author} {\bibfnamefont {Z.-B.}\ \bibnamefont {Zhu}}, \
  and\ \bibinfo {author} {\bibfnamefont {X.-G.}\ \bibnamefont {Huang}},\ }\href
  {\doibase 10.1103/PhysRevD.108.054006} {\bibfield  {journal} {\bibinfo
  {journal} {Phys. Rev. D}\ }\textbf {\bibinfo {volume} {108}},\ \bibinfo
  {pages} {054006} (\bibinfo {year} {2023})},\ \Eprint
  {http://arxiv.org/abs/2306.08362} {arXiv:2306.08362 [hep-ph]} \BibitemShut
  {NoStop}%
\bibitem [{\citenamefont {Wan}\ and\ \citenamefont
  {Ruggieri}(2021)}]{Wan:2020ffv}%
  \BibitemOpen
  \bibfield  {author} {\bibinfo {author} {\bibfnamefont {S.-S.}\ \bibnamefont
  {Wan}}\ and\ \bibinfo {author} {\bibfnamefont {M.}~\bibnamefont {Ruggieri}},\
  }\href {\doibase 10.1103/PhysRevD.103.094030} {\bibfield  {journal} {\bibinfo
   {journal} {Phys. Rev. D}\ }\textbf {\bibinfo {volume} {103}},\ \bibinfo
  {pages} {094030} (\bibinfo {year} {2021})},\ \Eprint
  {http://arxiv.org/abs/2012.12577} {arXiv:2012.12577 [hep-ph]} \BibitemShut
  {NoStop}%
\bibitem [{\citenamefont {Chen}\ \emph
  {et~al.}(2021{\natexlab{b}})\citenamefont {Chen}, \citenamefont {Zhang},
  \citenamefont {Li}, \citenamefont {Hou},\ and\ \citenamefont
  {Huang}}]{Chen:2020ath}%
  \BibitemOpen
  \bibfield  {author} {\bibinfo {author} {\bibfnamefont {X.}~\bibnamefont
  {Chen}}, \bibinfo {author} {\bibfnamefont {L.}~\bibnamefont {Zhang}},
  \bibinfo {author} {\bibfnamefont {D.}~\bibnamefont {Li}}, \bibinfo {author}
  {\bibfnamefont {D.}~\bibnamefont {Hou}}, \ and\ \bibinfo {author}
  {\bibfnamefont {M.}~\bibnamefont {Huang}},\ }\href {\doibase
  10.1007/JHEP07(2021)132} {\bibfield  {journal} {\bibinfo  {journal} {JHEP}\
  }\textbf {\bibinfo {volume} {07}},\ \bibinfo {pages} {132} (\bibinfo {year}
  {2021}{\natexlab{b}})},\ \Eprint {http://arxiv.org/abs/2010.14478}
  {arXiv:2010.14478 [hep-ph]} \BibitemShut {NoStop}%
\bibitem [{\citenamefont {Chen}\ \emph
  {et~al.}(2024{\natexlab{c}})\citenamefont {Chen}, \citenamefont {Chen},
  \citenamefont {Li},\ and\ \citenamefont {Huang}}]{Chen:2024jet}%
  \BibitemOpen
  \bibfield  {author} {\bibinfo {author} {\bibfnamefont {Y.}~\bibnamefont
  {Chen}}, \bibinfo {author} {\bibfnamefont {X.}~\bibnamefont {Chen}}, \bibinfo
  {author} {\bibfnamefont {D.}~\bibnamefont {Li}}, \ and\ \bibinfo {author}
  {\bibfnamefont {M.}~\bibnamefont {Huang}},\ }\href@noop {} {\  (\bibinfo
  {year} {2024}{\natexlab{c}})},\ \Eprint {http://arxiv.org/abs/2405.06386}
  {arXiv:2405.06386 [hep-ph]} \BibitemShut {NoStop}%
\bibitem [{\citenamefont {Braga}\ and\ \citenamefont
  {Junqueira}(2024)}]{Braga:2023qej}%
  \BibitemOpen
  \bibfield  {author} {\bibinfo {author} {\bibfnamefont {N.~R.~F.}\
  \bibnamefont {Braga}}\ and\ \bibinfo {author} {\bibfnamefont {O.~C.}\
  \bibnamefont {Junqueira}},\ }\href {\doibase 10.1016/j.physletb.2023.138330}
  {\bibfield  {journal} {\bibinfo  {journal} {Phys. Lett. B}\ }\textbf
  {\bibinfo {volume} {848}},\ \bibinfo {pages} {138330} (\bibinfo {year}
  {2024})},\ \Eprint {http://arxiv.org/abs/2306.08653} {arXiv:2306.08653
  [hep-th]} \BibitemShut {NoStop}%
\bibitem [{\citenamefont {Braga}\ \emph {et~al.}(2023)\citenamefont {Braga},
  \citenamefont {Ferreira},\ and\ \citenamefont {Junqueira}}]{Braga:2023qee}%
  \BibitemOpen
  \bibfield  {author} {\bibinfo {author} {\bibfnamefont {N.~R.~F.}\
  \bibnamefont {Braga}}, \bibinfo {author} {\bibfnamefont {L.~F.}\ \bibnamefont
  {Ferreira}}, \ and\ \bibinfo {author} {\bibfnamefont {O.~C.}\ \bibnamefont
  {Junqueira}},\ }\href {\doibase 10.1016/j.physletb.2023.138265} {\bibfield
  {journal} {\bibinfo  {journal} {Phys. Lett. B}\ }\textbf {\bibinfo {volume}
  {847}},\ \bibinfo {pages} {138265} (\bibinfo {year} {2023})},\ \Eprint
  {http://arxiv.org/abs/2301.01322} {arXiv:2301.01322 [hep-th]} \BibitemShut
  {NoStop}%
\bibitem [{\citenamefont {Zhao}\ \emph {et~al.}(2023)\citenamefont {Zhao},
  \citenamefont {He}, \citenamefont {Hou}, \citenamefont {Li},\ and\
  \citenamefont {Li}}]{Zhao:2022uxc}%
  \BibitemOpen
  \bibfield  {author} {\bibinfo {author} {\bibfnamefont {Y.-Q.}\ \bibnamefont
  {Zhao}}, \bibinfo {author} {\bibfnamefont {S.}~\bibnamefont {He}}, \bibinfo
  {author} {\bibfnamefont {D.}~\bibnamefont {Hou}}, \bibinfo {author}
  {\bibfnamefont {L.}~\bibnamefont {Li}}, \ and\ \bibinfo {author}
  {\bibfnamefont {Z.}~\bibnamefont {Li}},\ }\href {\doibase
  10.1007/JHEP04(2023)115} {\bibfield  {journal} {\bibinfo  {journal} {JHEP}\
  }\textbf {\bibinfo {volume} {04}},\ \bibinfo {pages} {115} (\bibinfo {year}
  {2023})},\ \Eprint {http://arxiv.org/abs/2212.14662} {arXiv:2212.14662
  [hep-ph]} \BibitemShut {NoStop}%
\bibitem [{\citenamefont {Yadav}(2023)}]{Yadav:2022qcl}%
  \BibitemOpen
  \bibfield  {author} {\bibinfo {author} {\bibfnamefont {G.}~\bibnamefont
  {Yadav}},\ }\href {\doibase 10.1016/j.physletb.2023.137925} {\bibfield
  {journal} {\bibinfo  {journal} {Phys. Lett. B}\ }\textbf {\bibinfo {volume}
  {841}},\ \bibinfo {pages} {137925} (\bibinfo {year} {2023})},\ \Eprint
  {http://arxiv.org/abs/2203.11959} {arXiv:2203.11959 [hep-th]} \BibitemShut
  {NoStop}%
\bibitem [{\citenamefont {Ebihara}\ \emph {et~al.}(2017)\citenamefont
  {Ebihara}, \citenamefont {Fukushima},\ and\ \citenamefont
  {Mameda}}]{Ebihara:2016fwa}%
  \BibitemOpen
  \bibfield  {author} {\bibinfo {author} {\bibfnamefont {S.}~\bibnamefont
  {Ebihara}}, \bibinfo {author} {\bibfnamefont {K.}~\bibnamefont {Fukushima}},
  \ and\ \bibinfo {author} {\bibfnamefont {K.}~\bibnamefont {Mameda}},\ }\href
  {\doibase 10.1016/j.physletb.2016.11.010} {\bibfield  {journal} {\bibinfo
  {journal} {Phys. Lett. B}\ }\textbf {\bibinfo {volume} {764}},\ \bibinfo
  {pages} {94} (\bibinfo {year} {2017})},\ \Eprint
  {http://arxiv.org/abs/1608.00336} {arXiv:1608.00336 [hep-ph]} \BibitemShut
  {NoStop}%
\bibitem [{\citenamefont {Montenegro}\ \emph {et~al.}(2017)\citenamefont
  {Montenegro}, \citenamefont {Tinti},\ and\ \citenamefont
  {Torrieri}}]{Montenegro:2017rbu}%
  \BibitemOpen
  \bibfield  {author} {\bibinfo {author} {\bibfnamefont {D.}~\bibnamefont
  {Montenegro}}, \bibinfo {author} {\bibfnamefont {L.}~\bibnamefont {Tinti}}, \
  and\ \bibinfo {author} {\bibfnamefont {G.}~\bibnamefont {Torrieri}},\ }\href
  {\doibase 10.1103/PhysRevD.96.056012} {\bibfield  {journal} {\bibinfo
  {journal} {Phys. Rev. D}\ }\textbf {\bibinfo {volume} {96}},\ \bibinfo
  {pages} {056012} (\bibinfo {year} {2017})},\ \bibinfo {note} {[Addendum:
  Phys.Rev.D 96, 079901 (2017)]},\ \Eprint {http://arxiv.org/abs/1701.08263}
  {arXiv:1701.08263 [hep-th]} \BibitemShut {NoStop}%
\bibitem [{\citenamefont {Florkowski}\ \emph {et~al.}(2018)\citenamefont
  {Florkowski}, \citenamefont {Friman}, \citenamefont {Jaiswal},\ and\
  \citenamefont {Speranza}}]{Florkowski:2017ruc}%
  \BibitemOpen
  \bibfield  {author} {\bibinfo {author} {\bibfnamefont {W.}~\bibnamefont
  {Florkowski}}, \bibinfo {author} {\bibfnamefont {B.}~\bibnamefont {Friman}},
  \bibinfo {author} {\bibfnamefont {A.}~\bibnamefont {Jaiswal}}, \ and\
  \bibinfo {author} {\bibfnamefont {E.}~\bibnamefont {Speranza}},\ }\href
  {\doibase 10.1103/PhysRevC.97.041901} {\bibfield  {journal} {\bibinfo
  {journal} {Phys. Rev. C}\ }\textbf {\bibinfo {volume} {97}},\ \bibinfo
  {pages} {041901} (\bibinfo {year} {2018})},\ \Eprint
  {http://arxiv.org/abs/1705.00587} {arXiv:1705.00587 [nucl-th]} \BibitemShut
  {NoStop}%
\bibitem [{\citenamefont {Braguta}\ \emph {et~al.}(2025)\citenamefont
  {Braguta}, \citenamefont {Chernodub},\ and\ \citenamefont
  {Roenko}}]{Braguta:2025ddq}%
  \BibitemOpen
  \bibfield  {author} {\bibinfo {author} {\bibfnamefont {V.~V.}\ \bibnamefont
  {Braguta}}, \bibinfo {author} {\bibfnamefont {M.~N.}\ \bibnamefont
  {Chernodub}}, \ and\ \bibinfo {author} {\bibfnamefont {A.~A.}\ \bibnamefont
  {Roenko}},\ }\href {\doibase 10.1103/xptn-qgfl} {\bibfield  {journal}
  {\bibinfo  {journal} {Phys. Rev. D}\ }\textbf {\bibinfo {volume} {111}},\
  \bibinfo {pages} {114508} (\bibinfo {year} {2025})},\ \Eprint
  {http://arxiv.org/abs/2503.18636} {arXiv:2503.18636 [hep-lat]} \BibitemShut
  {NoStop}%
\bibitem [{\citenamefont {Fukushima}\ and\ \citenamefont
  {Pu}(2021)}]{Fukushima:2020ucl}%
  \BibitemOpen
  \bibfield  {author} {\bibinfo {author} {\bibfnamefont {K.}~\bibnamefont
  {Fukushima}}\ and\ \bibinfo {author} {\bibfnamefont {S.}~\bibnamefont {Pu}},\
  }\href {\doibase 10.1016/j.physletb.2021.136346} {\bibfield  {journal}
  {\bibinfo  {journal} {Phys. Lett. B}\ }\textbf {\bibinfo {volume} {817}},\
  \bibinfo {pages} {136346} (\bibinfo {year} {2021})},\ \Eprint
  {http://arxiv.org/abs/2010.01608} {arXiv:2010.01608 [hep-th]} \BibitemShut
  {NoStop}%
\bibitem [{\citenamefont {Buzzegoli}\ and\ \citenamefont
  {Palermo}(2024)}]{Buzzegoli:2024mra}%
  \BibitemOpen
  \bibfield  {author} {\bibinfo {author} {\bibfnamefont {M.}~\bibnamefont
  {Buzzegoli}}\ and\ \bibinfo {author} {\bibfnamefont {A.}~\bibnamefont
  {Palermo}},\ }\href {\doibase 10.1103/PhysRevLett.133.262301} {\bibfield
  {journal} {\bibinfo  {journal} {Phys. Rev. Lett.}\ }\textbf {\bibinfo
  {volume} {133}},\ \bibinfo {pages} {262301} (\bibinfo {year} {2024})},\
  \Eprint {http://arxiv.org/abs/2407.14345} {arXiv:2407.14345 [hep-ph]}
  \BibitemShut {NoStop}%
\bibitem [{\citenamefont {Speranza}\ and\ \citenamefont
  {Weickgenannt}(2021)}]{Speranza:2020ilk}%
  \BibitemOpen
  \bibfield  {author} {\bibinfo {author} {\bibfnamefont {E.}~\bibnamefont
  {Speranza}}\ and\ \bibinfo {author} {\bibfnamefont {N.}~\bibnamefont
  {Weickgenannt}},\ }\href {\doibase 10.1140/epja/s10050-021-00455-2}
  {\bibfield  {journal} {\bibinfo  {journal} {Eur. Phys. J. A}\ }\textbf
  {\bibinfo {volume} {57}},\ \bibinfo {pages} {155} (\bibinfo {year} {2021})},\
  \Eprint {http://arxiv.org/abs/2007.00138} {arXiv:2007.00138 [nucl-th]}
  \BibitemShut {NoStop}%
\bibitem [{\citenamefont {Becattini}\ \emph {et~al.}(2019)\citenamefont
  {Becattini}, \citenamefont {Florkowski},\ and\ \citenamefont
  {Speranza}}]{Becattini:2018duy}%
  \BibitemOpen
  \bibfield  {author} {\bibinfo {author} {\bibfnamefont {F.}~\bibnamefont
  {Becattini}}, \bibinfo {author} {\bibfnamefont {W.}~\bibnamefont
  {Florkowski}}, \ and\ \bibinfo {author} {\bibfnamefont {E.}~\bibnamefont
  {Speranza}},\ }\href {\doibase 10.1016/j.physletb.2018.12.016} {\bibfield
  {journal} {\bibinfo  {journal} {Phys. Lett. B}\ }\textbf {\bibinfo {volume}
  {789}},\ \bibinfo {pages} {419} (\bibinfo {year} {2019})},\ \Eprint
  {http://arxiv.org/abs/1807.10994} {arXiv:1807.10994 [hep-th]} \BibitemShut
  {NoStop}%
\bibitem [{\citenamefont {Fang}\ \emph {et~al.}(2025)\citenamefont {Fang},
  \citenamefont {Fukushima}, \citenamefont {Pu},\ and\ \citenamefont
  {Wang}}]{Fang:2025aig}%
  \BibitemOpen
  \bibfield  {author} {\bibinfo {author} {\bibfnamefont {S.}~\bibnamefont
  {Fang}}, \bibinfo {author} {\bibfnamefont {K.}~\bibnamefont {Fukushima}},
  \bibinfo {author} {\bibfnamefont {S.}~\bibnamefont {Pu}}, \ and\ \bibinfo
  {author} {\bibfnamefont {D.-L.}\ \bibnamefont {Wang}},\ }\href@noop {} {\
  (\bibinfo {year} {2025})},\ \Eprint {http://arxiv.org/abs/2506.20698}
  {arXiv:2506.20698 [nucl-th]} \BibitemShut {NoStop}%
\bibitem [{\citenamefont {Huang}(2024)}]{Huang:2024ffg}%
  \BibitemOpen
  \bibfield  {author} {\bibinfo {author} {\bibfnamefont {X.-G.}\ \bibnamefont
  {Huang}},\ }\href@noop {} {\  (\bibinfo {year} {2024})},\ \Eprint
  {http://arxiv.org/abs/2411.11753} {arXiv:2411.11753 [nucl-th]} \BibitemShut
  {NoStop}%
\bibitem [{\citenamefont {Sakai}\ \emph {et~al.}(2010)\citenamefont {Sakai},
  \citenamefont {Sasaki}, \citenamefont {Kouno},\ and\ \citenamefont
  {Yahiro}}]{Sakai:2010rp}%
  \BibitemOpen
  \bibfield  {author} {\bibinfo {author} {\bibfnamefont {Y.}~\bibnamefont
  {Sakai}}, \bibinfo {author} {\bibfnamefont {T.}~\bibnamefont {Sasaki}},
  \bibinfo {author} {\bibfnamefont {H.}~\bibnamefont {Kouno}}, \ and\ \bibinfo
  {author} {\bibfnamefont {M.}~\bibnamefont {Yahiro}},\ }\href {\doibase
  10.1103/PhysRevD.82.076003} {\bibfield  {journal} {\bibinfo  {journal} {Phys.
  Rev. D}\ }\textbf {\bibinfo {volume} {82}},\ \bibinfo {pages} {076003}
  (\bibinfo {year} {2010})},\ \Eprint {http://arxiv.org/abs/1006.3648}
  {arXiv:1006.3648 [hep-ph]} \BibitemShut {NoStop}%
\bibitem [{\citenamefont {Sakai}\ \emph {et~al.}(2009)\citenamefont {Sakai},
  \citenamefont {Kashiwa}, \citenamefont {Kouno}, \citenamefont {Matsuzaki},\
  and\ \citenamefont {Yahiro}}]{Sakai:2009dv}%
  \BibitemOpen
  \bibfield  {author} {\bibinfo {author} {\bibfnamefont {Y.}~\bibnamefont
  {Sakai}}, \bibinfo {author} {\bibfnamefont {K.}~\bibnamefont {Kashiwa}},
  \bibinfo {author} {\bibfnamefont {H.}~\bibnamefont {Kouno}}, \bibinfo
  {author} {\bibfnamefont {M.}~\bibnamefont {Matsuzaki}}, \ and\ \bibinfo
  {author} {\bibfnamefont {M.}~\bibnamefont {Yahiro}},\ }\href {\doibase
  10.1103/PhysRevD.79.096001} {\bibfield  {journal} {\bibinfo  {journal} {Phys.
  Rev. D}\ }\textbf {\bibinfo {volume} {79}},\ \bibinfo {pages} {096001}
  (\bibinfo {year} {2009})},\ \Eprint {http://arxiv.org/abs/0902.0487}
  {arXiv:0902.0487 [hep-ph]} \BibitemShut {NoStop}%
\bibitem [{\citenamefont {Yamamoto}(2011)}]{Yamamoto:2011gk}%
  \BibitemOpen
  \bibfield  {author} {\bibinfo {author} {\bibfnamefont {A.}~\bibnamefont
  {Yamamoto}},\ }\href {\doibase 10.1103/PhysRevLett.107.031601} {\bibfield
  {journal} {\bibinfo  {journal} {Phys. Rev. Lett.}\ }\textbf {\bibinfo
  {volume} {107}},\ \bibinfo {pages} {031601} (\bibinfo {year} {2011})},\
  \Eprint {http://arxiv.org/abs/1105.0385} {arXiv:1105.0385 [hep-lat]}
  \BibitemShut {NoStop}%
\bibitem [{\citenamefont {Ambrus}\ and\ \citenamefont
  {Chernodub}(2023)}]{Ambrus:2019khr}%
  \BibitemOpen
  \bibfield  {author} {\bibinfo {author} {\bibfnamefont {V.~E.}\ \bibnamefont
  {Ambrus}}\ and\ \bibinfo {author} {\bibfnamefont {M.~N.}\ \bibnamefont
  {Chernodub}},\ }\href {\doibase 10.1140/epjc/s10052-023-11244-0} {\bibfield
  {journal} {\bibinfo  {journal} {Eur. Phys. J. C}\ }\textbf {\bibinfo {volume}
  {83}},\ \bibinfo {pages} {111} (\bibinfo {year} {2023})},\ \bibinfo {note}
  {[Erratum: Eur.Phys.J.C 84, 289 (2024)]},\ \Eprint
  {http://arxiv.org/abs/1912.11034} {arXiv:1912.11034 [hep-th]} \BibitemShut
  {NoStop}%
\bibitem [{\citenamefont {Chernodub}\ and\ \citenamefont
  {Ambrus}(2021)}]{Chernodub:2020yaf}%
  \BibitemOpen
  \bibfield  {author} {\bibinfo {author} {\bibfnamefont {M.~N.}\ \bibnamefont
  {Chernodub}}\ and\ \bibinfo {author} {\bibfnamefont {V.~E.}\ \bibnamefont
  {Ambrus}},\ }\href {\doibase 10.1103/PhysRevD.103.094015} {\bibfield
  {journal} {\bibinfo  {journal} {Phys. Rev. D}\ }\textbf {\bibinfo {volume}
  {103}},\ \bibinfo {pages} {094015} (\bibinfo {year} {2021})},\ \Eprint
  {http://arxiv.org/abs/2005.03575} {arXiv:2005.03575 [hep-th]} \BibitemShut
  {NoStop}%
\bibitem [{\citenamefont {Brandt}\ \emph {et~al.}(2024)\citenamefont {Brandt},
  \citenamefont {Endr{\H{o}}di}, \citenamefont {Garnacho-Velasco},\ and\
  \citenamefont {Mark{\'o}}}]{Brandt:2024wlw}%
  \BibitemOpen
  \bibfield  {author} {\bibinfo {author} {\bibfnamefont {B.~B.}\ \bibnamefont
  {Brandt}}, \bibinfo {author} {\bibfnamefont {G.}~\bibnamefont
  {Endr{\H{o}}di}}, \bibinfo {author} {\bibfnamefont {E.}~\bibnamefont
  {Garnacho-Velasco}}, \ and\ \bibinfo {author} {\bibfnamefont
  {G.}~\bibnamefont {Mark{\'o}}},\ }\href {\doibase 10.1007/JHEP09(2024)092}
  {\bibfield  {journal} {\bibinfo  {journal} {JHEP}\ }\textbf {\bibinfo
  {volume} {09}},\ \bibinfo {pages} {092} (\bibinfo {year} {2024})},\ \Eprint
  {http://arxiv.org/abs/2405.09484} {arXiv:2405.09484 [hep-lat]} \BibitemShut
  {NoStop}%
\bibitem [{\citenamefont {Ratti}\ \emph {et~al.}(2006)\citenamefont {Ratti},
  \citenamefont {Thaler},\ and\ \citenamefont {Weise}}]{Ratti:2005jh}%
  \BibitemOpen
  \bibfield  {author} {\bibinfo {author} {\bibfnamefont {C.}~\bibnamefont
  {Ratti}}, \bibinfo {author} {\bibfnamefont {M.~A.}\ \bibnamefont {Thaler}}, \
  and\ \bibinfo {author} {\bibfnamefont {W.}~\bibnamefont {Weise}},\ }\href
  {\doibase 10.1103/PhysRevD.73.014019} {\bibfield  {journal} {\bibinfo
  {journal} {Phys. Rev. D}\ }\textbf {\bibinfo {volume} {73}},\ \bibinfo
  {pages} {014019} (\bibinfo {year} {2006})},\ \Eprint
  {http://arxiv.org/abs/hep-ph/0506234} {arXiv:hep-ph/0506234} \BibitemShut
  {NoStop}%
\bibitem [{\citenamefont {Ratti}\ \emph {et~al.}(2007)\citenamefont {Ratti},
  \citenamefont {Roessner}, \citenamefont {Thaler},\ and\ \citenamefont
  {Weise}}]{Ratti:2006wg}%
  \BibitemOpen
  \bibfield  {author} {\bibinfo {author} {\bibfnamefont {C.}~\bibnamefont
  {Ratti}}, \bibinfo {author} {\bibfnamefont {S.}~\bibnamefont {Roessner}},
  \bibinfo {author} {\bibfnamefont {M.~A.}\ \bibnamefont {Thaler}}, \ and\
  \bibinfo {author} {\bibfnamefont {W.}~\bibnamefont {Weise}},\ }\href
  {\doibase 10.1140/epjc/s10052-006-0065-x} {\bibfield  {journal} {\bibinfo
  {journal} {Eur. Phys. J. C}\ }\textbf {\bibinfo {volume} {49}},\ \bibinfo
  {pages} {213} (\bibinfo {year} {2007})},\ \Eprint
  {http://arxiv.org/abs/hep-ph/0609218} {arXiv:hep-ph/0609218} \BibitemShut
  {NoStop}%
\bibitem [{\citenamefont {Fukushima}\ and\ \citenamefont
  {Sasaki}(2013)}]{Fukushima:2013rx}%
  \BibitemOpen
  \bibfield  {author} {\bibinfo {author} {\bibfnamefont {K.}~\bibnamefont
  {Fukushima}}\ and\ \bibinfo {author} {\bibfnamefont {C.}~\bibnamefont
  {Sasaki}},\ }\href {\doibase 10.1016/j.ppnp.2013.05.003} {\bibfield
  {journal} {\bibinfo  {journal} {Prog. Part. Nucl. Phys.}\ }\textbf {\bibinfo
  {volume} {72}},\ \bibinfo {pages} {99} (\bibinfo {year} {2013})},\ \Eprint
  {http://arxiv.org/abs/1301.6377} {arXiv:1301.6377 [hep-ph]} \BibitemShut
  {NoStop}%
\bibitem [{\citenamefont {Hansen}\ \emph {et~al.}(2007)\citenamefont {Hansen},
  \citenamefont {Alberico}, \citenamefont {Beraudo}, \citenamefont {Molinari},
  \citenamefont {Nardi},\ and\ \citenamefont {Ratti}}]{Hansen:2006ee}%
  \BibitemOpen
  \bibfield  {author} {\bibinfo {author} {\bibfnamefont {H.}~\bibnamefont
  {Hansen}}, \bibinfo {author} {\bibfnamefont {W.~M.}\ \bibnamefont
  {Alberico}}, \bibinfo {author} {\bibfnamefont {A.}~\bibnamefont {Beraudo}},
  \bibinfo {author} {\bibfnamefont {A.}~\bibnamefont {Molinari}}, \bibinfo
  {author} {\bibfnamefont {M.}~\bibnamefont {Nardi}}, \ and\ \bibinfo {author}
  {\bibfnamefont {C.}~\bibnamefont {Ratti}},\ }\href {\doibase
  10.1103/PhysRevD.75.065004} {\bibfield  {journal} {\bibinfo  {journal} {Phys.
  Rev. D}\ }\textbf {\bibinfo {volume} {75}},\ \bibinfo {pages} {065004}
  (\bibinfo {year} {2007})},\ \Eprint {http://arxiv.org/abs/hep-ph/0609116}
  {arXiv:hep-ph/0609116} \BibitemShut {NoStop}%
\bibitem [{\citenamefont {Buballa}(2005)}]{Buballa:2003qv}%
  \BibitemOpen
  \bibfield  {author} {\bibinfo {author} {\bibfnamefont {M.}~\bibnamefont
  {Buballa}},\ }\href {\doibase 10.1016/j.physrep.2004.11.004} {\bibfield
  {journal} {\bibinfo  {journal} {Phys. Rept.}\ }\textbf {\bibinfo {volume}
  {407}},\ \bibinfo {pages} {205} (\bibinfo {year} {2005})},\ \Eprint
  {http://arxiv.org/abs/hep-ph/0402234} {arXiv:hep-ph/0402234} \BibitemShut
  {NoStop}%
\bibitem [{\citenamefont {Endr\H{o}di}\ and\ \citenamefont
  {Mark\'o}(2019)}]{Endrodi:2019whh}%
  \BibitemOpen
  \bibfield  {author} {\bibinfo {author} {\bibfnamefont {G.}~\bibnamefont
  {Endr\H{o}di}}\ and\ \bibinfo {author} {\bibfnamefont {G.}~\bibnamefont
  {Mark\'o}},\ }\href {\doibase 10.1007/JHEP08(2019)036} {\bibfield  {journal}
  {\bibinfo  {journal} {JHEP}\ }\textbf {\bibinfo {volume} {08}},\ \bibinfo
  {pages} {036} (\bibinfo {year} {2019})},\ \Eprint
  {http://arxiv.org/abs/1905.02103} {arXiv:1905.02103 [hep-lat]} \BibitemShut
  {NoStop}%
\bibitem [{\citenamefont {Schaefer}\ \emph {et~al.}(2007)\citenamefont
  {Schaefer}, \citenamefont {Pawlowski},\ and\ \citenamefont
  {Wambach}}]{Schaefer:2007pw}%
  \BibitemOpen
  \bibfield  {author} {\bibinfo {author} {\bibfnamefont {B.-J.}\ \bibnamefont
  {Schaefer}}, \bibinfo {author} {\bibfnamefont {J.~M.}\ \bibnamefont
  {Pawlowski}}, \ and\ \bibinfo {author} {\bibfnamefont {J.}~\bibnamefont
  {Wambach}},\ }\href {\doibase 10.1103/PhysRevD.76.074023} {\bibfield
  {journal} {\bibinfo  {journal} {Phys. Rev. D}\ }\textbf {\bibinfo {volume}
  {76}},\ \bibinfo {pages} {074023} (\bibinfo {year} {2007})},\ \Eprint
  {http://arxiv.org/abs/0704.3234} {arXiv:0704.3234 [hep-ph]} \BibitemShut
  {NoStop}%
\bibitem [{\citenamefont {Tan}\ \emph {et~al.}(2022)\citenamefont {Tan},
  \citenamefont {Dore}, \citenamefont {Dexheimer}, \citenamefont
  {Noronha-Hostler},\ and\ \citenamefont {Yunes}}]{PhysRevD.105.023018}%
  \BibitemOpen
  \bibfield  {author} {\bibinfo {author} {\bibfnamefont {H.}~\bibnamefont
  {Tan}}, \bibinfo {author} {\bibfnamefont {T.}~\bibnamefont {Dore}}, \bibinfo
  {author} {\bibfnamefont {V.}~\bibnamefont {Dexheimer}}, \bibinfo {author}
  {\bibfnamefont {J.}~\bibnamefont {Noronha-Hostler}}, \ and\ \bibinfo {author}
  {\bibfnamefont {N.}~\bibnamefont {Yunes}},\ }\href {\doibase
  10.1103/PhysRevD.105.023018} {\bibfield  {journal} {\bibinfo  {journal}
  {Phys. Rev. D}\ }\textbf {\bibinfo {volume} {105}},\ \bibinfo {pages}
  {023018} (\bibinfo {year} {2022})}\BibitemShut {NoStop}%
\bibitem [{\citenamefont {Pinto}(2023)}]{PhysRevC.107.045807}%
  \BibitemOpen
  \bibfield  {author} {\bibinfo {author} {\bibfnamefont {M.~B.}\ \bibnamefont
  {Pinto}},\ }\href {\doibase 10.1103/PhysRevC.107.045807} {\bibfield
  {journal} {\bibinfo  {journal} {Phys. Rev. C}\ }\textbf {\bibinfo {volume}
  {107}},\ \bibinfo {pages} {045807} (\bibinfo {year} {2023})}\BibitemShut
  {NoStop}%
\bibitem [{\citenamefont {Kneur}\ \emph {et~al.}(2010)\citenamefont {Kneur},
  \citenamefont {Pinto},\ and\ \citenamefont {Ramos}}]{PhysRevC.81.065205}%
  \BibitemOpen
  \bibfield  {author} {\bibinfo {author} {\bibfnamefont {J.-L.}\ \bibnamefont
  {Kneur}}, \bibinfo {author} {\bibfnamefont {M.~B.}\ \bibnamefont {Pinto}}, \
  and\ \bibinfo {author} {\bibfnamefont {R.~O.}\ \bibnamefont {Ramos}},\ }\href
  {\doibase 10.1103/PhysRevC.81.065205} {\bibfield  {journal} {\bibinfo
  {journal} {Phys. Rev. C}\ }\textbf {\bibinfo {volume} {81}},\ \bibinfo
  {pages} {065205} (\bibinfo {year} {2010})}\BibitemShut {NoStop}%
\bibitem [{\citenamefont {Klevansky}(1992)}]{Klevansky:1992qe}%
  \BibitemOpen
  \bibfield  {author} {\bibinfo {author} {\bibfnamefont {S.~P.}\ \bibnamefont
  {Klevansky}},\ }\href {\doibase 10.1103/RevModPhys.64.649} {\bibfield
  {journal} {\bibinfo  {journal} {Rev. Mod. Phys.}\ }\textbf {\bibinfo {volume}
  {64}},\ \bibinfo {pages} {649} (\bibinfo {year} {1992})}\BibitemShut
  {NoStop}%
\end{thebibliography}%

\end{document}